\documentclass{article}
\usepackage{graphicx} 

\usepackage[utf8]{inputenc}

\usepackage[english]{babel}
\usepackage[
    backend=bibtex,      
    style=authoryear,    
    sorting=nyt
]{biblatex}
\usepackage{csquotes}

\usepackage{enotez}
\usepackage{authblk}

\usepackage{graphicx}
\usepackage[a4paper]{geometry}
\usepackage[pdfborder={0 0 0}]{hyperref}
\usepackage[utf8]{inputenc}
\usepackage[T1]{fontenc}
\usepackage{dsfont, amsmath, amsfonts, amssymb, amsthm, units}
\usepackage{enumerate}
\usepackage{tikz-cd}
\usepackage{bbold}
\usepackage{hyperref,bookmark}
\usepackage{slashed}
\usepackage{mathtools}
\usepackage{palatino}
\usepackage{euler}
\usepackage{verbatim}

\DeclarePairedDelimiterX\braket[2]{\langle}{\rangle}{#1\,\delimsize\vert\,\mathopen{}#2}

\newcommand{\R}{\mathbb{R}}
\newcommand{\C}{\mathbb{C}}

\theoremstyle{definition}
\newtheorem{definition}{Definition}[section]

\newtheorem{theorem}[definition]{Theorem}

\newtheorem{proposition}[definition]{Proposition}

\title{Global Gauge Symmetry Breaking in the Abelian Higgs Mechanism} 
\author[1,*]{Silvester G.A. Borsboom}
\author[2]{Sebastian De Haro}
\affil[1]{\normalsize Radboud Center for Natural Philosophy and Institute for Mathematics, Astrophysics and Particle Physics, Radboud University Nijmegen}
\affil[2]{\normalsize Institute for Logic, Language and Computation and Institute of Physics, University of Amsterdam}
\affil[*]{Corresponding author: silvester.borsboom@ru.nl}
\date{\today}

\begin{document}

\maketitle

\begin{abstract}

This paper aims to resolve the incompatibility between two extant gauge-invariant accounts of the Abelian Higgs mechanism: the first account uses global gauge symmetry breaking, and the second {\it eliminates} spontaneous symmetry breaking entirely. We resolve this incompatibility by using the constrained Hamiltonian formalism in symplectic geometry. First we argue that, unlike their local counterparts, global gauge symmetries are physical in the presence of boundary conditions. The symmetry that is spontaneously broken by the Higgs mechanism is this global one. Second, we explain how the Coulomb gauge is the preferred gauge for a gauge-invariant account of the Abelian Higgs mechanism. Based on the existence of the physical global gauge symmetry, we resolve the incompatibility between the two accounts by arguing that the correct way to carry out the second method is to eliminate only the redundant gauge symmetries, i.e.~those local gauge symmetries which are not global. We extend our analysis to quantum field theory, where we show that the Abelian Higgs mechanism can be understood as spontaneous global $U(1)$ symmetry breaking in the $C^*$-algebraic sense.\endnote{This article is based on the master thesis of the corresponding author, which was supervised by the other author and by Hessel Posthuma. See \autocite{borsboom2024spontaneous}.}
\end{abstract}

\thispagestyle{empty}
\clearpage
\thispagestyle{empty}

\begin{quote}
\tableofcontents
\end{quote}

\thispagestyle{empty}
\clearpage
\setcounter{page}{1}

\section{Introduction}\label{intro}

In the original physics literature, the Higgs mechanism is presented as an instance of spontaneous local gauge symmetry breaking \autocite{Higgs:1964,higgsBrokenSymmetriesMasses1964,englertBrokenSymmetryMass1964,guralnikGlobalConservationLaws1964}, similar to symmetry breaking in the Ginzburg-Landau (GL) theory of superconductivity \autocite{Ginzburg:1950sr,ginzburgTheorySuperconductivity2009}. 

But over the past two decades, philosophers of physics have pointed out the incongruity of the notion of ``local gauge symmetry breaking,'' and they have consequently pleaded for a gauge-invariant account of the Higgs mechanism. In this literature, the main conceptual problem that is levelled against the standard narrative seems to stem from the idea that ``a genuine property like mass cannot be gained by eating descriptive fluff, which is just what gauge
is'' \autocite{earmanLawsSymmetrySymmetry2004}. This problem is further exacerbated by Elitzur's theorem \autocite{elitzurImpossibilitySpontaneouslyBreaking1975,smeenkElusiveHiggsMechanism2006}, which forbids local gauge symmetry breaking by showing that gauge-dependent quantities must have a vanishing vacuum expectation value (VEV). This applies in particular to the Higgs field, which supposedly acquires a nonzero VEV in the electroweak phase transition.\\
\\
{\it Two incompatible proposals.} In response to the problems outlined above, philosophers have sought to reformulate the Higgs mechanism in a way that avoids the dubious concept of `local gauge symmetry breaking'. For example, Lyre has argued that the Higgs mechanism does not exist at all, because it is a mere rewriting of Lagrangians using different choices of gauge \autocite{lyreDoesHiggsMechanism2008}.
His argument comes down to the idea that, unlike symmetry transformations in other cases of spontaneous symmetry breaking (SSB), gauge transformations do not correspond to anything physical. Thus, according to Lyre, the Higgs mechanism is only a heuristic with no explanatory value. 

In this paper, we will be concerned with two incompatible accounts of the Higgs mechanism. While the first account acknowledges SSB, the second 
denies that a gauge symmetry can be spontaneously broken (and in this particular aspect it agrees with Lyre), as follows:\\
\\
(i) {\it The Higgs mechanism as the spontaneous breaking of a global gauge symmetry.} Building on a definition of `gauge' as leading to  a breakdown of determinism (as opposed to the definition of `gauge' as simply all spacetime-dependent $U(1)$ transformations), the option pursued in Section 7.2 of \autocite{struyveGaugeInvariantAccounts2011a} is to formulate the Abelian Higgs mechanism in a substantive gauge-invariant way, in terms of the spontaneous breaking of a \textit{global} (also known as \textit{rigid}), rather than a local, gauge symmetry. In this approach, one redefines the Hamiltonian of the Abelian Higgs model in terms of the transverse component of the electromagnetic potential, which is gauge-invariant, and a dressed version of the Higgs field, which is invariant under all gauge transformations except the global ones. This Hamiltonian contains a massless gauge field and exhibits a residual global $U(1)$ gauge symmetry. When this global symmetry is broken in the usual way, an expansion up to quadratic order around the chosen vacuum configuration yields a Hamiltonian with a massive gauge field. By thus breaking the {\it global} gauge symmetry, this account succeeds in producing massive gauge fields. This is also in line with the Josephson effect \autocite{josephsonPossibleNewEffects1962,vanwezelSpontaneousSymmetryBreaking2008}, in which a current flows between two superconductors depending on the relative \textit{global} phase of the symmetry-breaking wavefunction of Cooper pairs.\endnote{For a much more in-depth study of the relational understanding of symmetry breaking in superconductors, see \autocite{borsboomdupont}. See also Chapter 6 \autocite{DeHaro:2025gby} for the link between superconductors and the Higgs mechanism by means of duality.} Thus, differences in this global $U(1)$ gauge phase can be measured across a boundary between subsystem and environment, in accordance with the relational interpretation of symmetries.

Struyve stresses that global and local gauge symmetries are to be treated differently only in the presence of \textit{boundary conditions}. If there are no boundary conditions, then both local and global gauge transformations lead to a breakdown of determinism in the same way, and both should be considered unphysical according to this definition of `gauge'. This raises the following question: why should a reformulation of the Higgs mechanism in terms of fields that are invariant under local, but not global, gauge transformations be called gauge-invariant? Does this not suffer from the same conceptual issues pointed out in \autocite{lyreDoesHiggsMechanism2008}? In fact, \autocite{friederichGaugeSymmetryBreaking2013} discusses the spontaneous breaking of various remnant global gauge symmetries and points out that they do not match up with actual phase transitions, rendering it difficult to interpret the Higgs mechanism in that way. Thus it may be more promising to attempt to rid the Higgs mechanism of SSB altogether. \\
\\
(ii) {\it The Higgs field sliding down a potential: no SSB.} This approach is advocated by \autocite{berghoferGaugeSymmetriesSymmetry2023},\endnote{See also \autocite{massonRemarkSpontaneousSymmetry2010,fournelGaugeInvariantComposite2014,francoisArtificialSubstantialGauge2019,attardDressingFieldMethod2018,zajacDressingFieldMethod2023}.} and was also discussed in Section 4.2 of \autocite{struyveGaugeInvariantAccounts2011a} and originally by Higgs himself \autocite{higgsSpontaneousSymmetryBreakdown1966}.
Berghofer et al. use the dressing field method (DFM), with a dressing field built from the polar decomposition of the Higgs field, to rewrite the Abelian Higgs and electroweak Lagrangians in terms of {\it gauge-invariant dressed fields}. Since the dressed fields are gauge-invariant, there can be no gauge symmetry breaking, and the Higgs mechanism must be interpreted by way of the Higgs field sliding down its potential $V\colon\R^+\to\R$, which in this approach only takes strictly positive values as its input (i.e.~the Higgs field is not permitted to vanish anywhere: and recall that the Higgs potential only depends on the modulus of the Higgs field). These ideas can be generalized to perturbative quantum field theory (QFT) through the Fröhlich-Morchio-Strocchi (FMS) approach \autocite{frohlichHiggsPhenomenonSymmetry1980,frohlichHiggsPhenomenonSymmetry1981}, in which $n$-point functions of gauge-invariant composite fields are expressed in terms of $n$-point functions of the elementary fields. It can then be shown that quantities like gauge boson masses agree on either side of this expression, i.e.~for the gauge-invariant composite fields and gauge-dependent elementary fields \autocite{maasBroutEnglertHiggs2019,maasFrohlichMorchioStrocchiMechanismUnderestimated2023}.\\

These two gauge-invariant accounts of the Higgs mechanism are clearly incompatible: (i) the approach of \autocite[Section 7.2]{struyveGaugeInvariantAccounts2011a} reformulates the Abelian Higgs mechanism in the Coulomb gauge in terms of fields that are invariant under all the gauge symmetries except the global ones, which are {\it spontaneously broken}. (ii) Berghofer et al., in the spirit of \autocite[Section 4.2]{struyveGaugeInvariantAccounts2011a} and \autocite{higgsSpontaneousSymmetryBreakdown1966}, advocate the idea that, by using the DFM and FMS approach, SSB should be {\it eliminated} from the Higgs mechanism altogether.

The above suggests our central question: what does it mean to give a ``gauge-invariant account'' of the Higgs mechanism? Should such an account include, or remove, global gauge symmetry, i.e.~should we dress our elementary fields in order to turn them into entirely gauge-invariant composite objects, or should we leave room for global symmetries?  Or in the language of \autocite{struyveGaugeInvariantAccounts2011a}: should one take `gauge' to designate all spacetime-dependent $U(1)$ gauge transformations or only those that lead, after taking into account boundary conditions, to a breakdown of determinism?

In this paper, we will build upon recent results which show how global gauge symmetries arise as the physical symmetry group of Yang-Mills theory on a Cauchy surface with asymptotic boundary conditions at spatial infinity \autocite{borsboom}. This means that we consider gauge transformations `at a time', rather than transformations of histories as in \autocite{struyveGaugeInvariantAccounts2011a}. Global gauge symmetries respect asymptotic boundary conditions on the Yang-Mills fields, but are not `trivial' or `redundant' because they are not generated by the Hamiltonian constraints of the theory. Physically, this can be seen from their non-trivial action on charged matter fields. In Yang-Mills-Higgs theory the same holds, but only in the \textit{unbroken} phase. In the \textit{broken} phase, on the other hand, the boundary-preserving and the trivial symmetries overlap (up to topological terms). These results suggest that gauge symmetry breaking in the Higgs mechanism should be understood as global gauge symmetry breaking.

Because of the initial presence of a physical global gauge symmetry, the DFM cannot be used at the level of the theory's finite-dimensional principal fibre bundle to remove all gauge symmetry, as is done also when working in unitary gauge. It must instead be used at the level of the infinite-dimensional configuration space of fields, a method pursued by means of the so-called \textit{field-dependent DFM} \autocite{fournelGaugeInvariantComposite2014,Francois:2020tom,Francois:2021jrk,berghoferDressingVsFixing2024,Francois:2024rdm}, as well as in work on boundaries in Yang-Mills theory \autocite{gomesUnifiedGeometricFramework2018,gomesGaugingBoundaryFieldspace2019,gomesQuasilocalDegreesFreedom2021} In fact, this corresponds---at least in the Abelian case---to projecting down to the Coulomb gauge. This treatment is \textit{relational}, in that it singles out the rigid global gauge symmetries as physical in the sense of Galileo's ship (i.e. when applying them to subsystems) \autocite{gomesHolismEmpiricalSignificance2021}. We will argue that the dressing field method, thus corrected, vindicates the gauge-invariant account (i) found in Section 7.2 of \autocite{struyveGaugeInvariantAccounts2011a}.

However, even if this incompatibility is resolved, the problem of the Higgs mechanism in QFT remains. Indeed, Stöltzner already expressed the worry that, even though Struyve provided a gauge-invariant account of the classical Higgs mechanism, the same remained to be done for the quantum case \autocite{stoltznerConstrainingHiggsMechanism2012}. This refers to Earman's original desire to relate SSB in the Higgs mechanism to the algebraic definition of SSB in quantum systems in terms of unitary inequivalence of GNS representations \autocite{earmanRoughGuideSpontaneous2003}. Although the quantization of gauge theories is poorly understood, the desired link can be made by considering global gauge symmetry breaking. This was proved in \autocite{morchioLocalizationSymmetries2007}, who showed that there can only be massive photons in the Abelian Higgs model if the global $U(1)$ symmetry of the theory is spontaneously broken in the $C^*$-algebraic sense. Conceptualizing this result is the second major aim of this article.

Thus the paper falls into two parts: first, Sections \ref{globalgaugeconstraint} to \ref{coulomb} discuss classical field theory. Second, Section \ref{higgsqft} discusses QFT. In the first part, Section \ref{globalgaugeconstraint} introduces preliminary terminology on the physical status of gauge symmetries. After considering the details of the Hamiltonian formulation of electromagnetism in Section \ref{electrogauss}, we turn, in Section \ref{coulomb}, to resolving the incompatibility between the two accounts, (i) and (ii). We explain how the Coulomb gauge can be implemented by means of a dressing field, i.e. as in ~(ii), so that the remnant subgroup is precisely the group of global gauge transformations, and we discuss how this is an agreement with account (i). Finally, in Section \ref{higgsqft}, we recall the definition of SSB in algebraic quantum theory and show how the Abelian Higgs mechanism can be formulated by means of global gauge symmetry breaking in that language.

\section{Global Gauge Symmetries as the Physical Gauge Group}\label{globalgaugeconstraint}

The aim of this Section is to explain how global gauge symmetries arise as the group of physical symmetries in Yang-Mills theory, and in particular electromagnetism, with boundary conditions. To this end, we begin with an overview of the philosophical terminology surrounding the physical status of gauge symmetries in Section \ref{gfgt}, focussing on the distinction between two conceptions of the term `gauge', one mathematical and the other interpretative. In Section \ref{sec:DESglobal} we then explain how global gauge symmetries can exhibit direct empirical significance. Finally, in Section \ref{gps} we introduce various subgroups of the gauge group and explain how a certain quotient of two of these subgroups should be viewed as the group of physical gauge symmetries, and how this quotient equals the global gauge group.

\subsection{Gauge fields, gauge transformations and redundancies}\label{gfgt}

This Section does two things: first, it recalls the basic mathematical notions that we will use, and how they relate to physicists' way of thinking about gauge transformations. Second, it introduces the relevant interpretative notions from the literature, and especially that of a {\it redundant} gauge symmetry.\\
\\
(1)~~{\it Formal description of a gauge theory.} In mathematical gauge theory, a {\it gauge field} is a connection on a principal $G$-bundle $P$ over a differentiable manifold $M$ which represents space(time). Such a bundle locally looks like a product space $U\times G$, where $U\subset M$, but its global structure can be more complicated. Here, $G$ is called the {\it structure group}, and it is a compact matrix Lie group that acts on the bundle $P$. It is a global group, i.e.~its generators do not vary over the manifold $M$. The bundle has a projection $\pi:P\to M$ onto the base manifold. In what follows, when working in the Hamiltonian formalism, we will work with a bundle that is defined only over space and not spacetime. In that case we denote the base space by $\Sigma$ instead of $M$. Points on a spatial three-manifold $\Sigma$ will be denoted by the boldface letter $\bf x$, while points on the spacetime four-manifold $M$ will be denoted by $x$.

A {\it gauge transformation}, $f$, is a bundle automorphism, $f:P\to P$. The group $\mathcal{G}(P)$ (or simply $\mathcal{G}$), of such automorphisms under composition, is called the \textit{gauge group}. One should clearly distinguish between the structure group $G$, which is global, and the gauge group $\mathcal{G}$, which is local, i.e.~such that its generators vary over the points of the base manifold $M$ (and is therefore infinite-dimensional).

Note that this is {\it not} the definition of a gauge transformation that is commonly used by physicists, who usually think of gauge transformations as $G$-valued maps on spacetime, i.e.~as {\it maps from the manifold $M$ into the structure group $G$}. However, it {\it is} possible to think of gauge transformations in this way, by using the following two isomorphisms: 

(i)~~The group isomorphism\endnote{It is not too hard to check that the isomorphism is provided by sending $f\in\mathcal{G}(P)$ to $\sigma_f\colon P\to G$, defined via $f(p)=p\sigma_f(p)$ for any $p\in P$.} $\mathcal{G}(P)\cong C^\infty(P,G)^G$, where $C^\infty(P,G)^G$ denotes the group of smooth maps $f\colon P\to G$ satisfying $f(ph)=h^{-1}ph$ for all $p\in P,h\in G$ (for more details and proofs, see Chapter 5 of \autocite{hamiltonMathematicalGaugeTheory2017}). By this isomorphism, gauge transformations can be viewed as {\it $G$-valued} maps on the principal bundle. 

These are not quite $G$-valued maps {\it on spacetime} yet, i.e.~we would like to work with $C^\infty(M,G)$ rather than $C^\infty(P,G)^G$, which contains the larger and more abstract space $P$. This will be possible, thanks to the second isomorphism, but only if we choose a (local) gauge. A (local) {\it gauge} is a (local) section of the bundle $P$, i.e.~a smooth map $s\colon U\to P$ that satisfies $\pi\circ s=\text{id}_U$, where $U\subset M$ is an open subset of the manifold. If this open subset is the whole manifold $M$, then the bundle is called \textit{trivializable}.\endnote{For examples of non-trivializable bundles, one looks for example at a manifold $M$ that is topologically non-trivial (where, by `topologically trivial' manifold, we mean one where all the homotopy groups are the identity). This is for example the case for BPST SU(2) instantons on $S^3$ (where `BPST' stands for the names of the authors of \autocite{belavin1975pseudoparticle}), which are characterized by the second Chern class of the principal bundle.}

(ii)~~It can be shown that a section of the bundle, $s$, defines an isomorphism\endnote{This isomorphism simply sends $f\in C^\infty(P_U,G)^G$ to $f\circ s\colon M\to G$.} of groups $C^\infty(P_U,G)^G\cong C^\infty(U,G)$, where $P_U$ is the restricted principal $G$-bundle over $U$. Thus, when we work with a trivializable principal bundle, we can extend $C^\infty(U,G)$ to $C^\infty(M,G)$.

In other words: through these two isomorphisms, gauge transformations can be viewed as $G$-valued maps on the manifold, but \textit{only in a specific gauge}. This allows us to define \textit{global gauge transformations} as the constant $G$-valued maps on (a region of) the manifold. We note that, if there does not exist a global section of the principal bundle (i.e.~if the bundle is not trivializable), then there is no well-defined notion of `global gauge transformation' on the whole manifold, but only locally. However, a `local global gauge transformation' seems like a \textit{contradictio in terminis}, as we would like to think of global gauge transformations as extending out to infinity, since they do not depend on spacetime coordinates. For this reason, some authors prefer the word \textit{rigid} over \textit{global}. (Since for us there will be no confusion, we will continue to say `global'.)

If one works on a {\it trivializable} principal bundle over a manifold $M$, the gauge group $\mathcal{G}$ will always be isomorphic to $C^\infty(M,G)$. Thus, whenever we work with a trivializable bundle in this article, we will not need to insist on the distinction between $\mathcal{G}$ and $C^\infty(M,G)$, and we will sometimes simply write: $\mathcal{G}=C^\infty(M,G)$.

Note that, on a trivializable principal bundle, the {\it subgroup of $\mathcal{G}$ that consists only of global gauge transformations} can be identified with (i.e.~is isomorphic to) the structure group $G$. However, it is only in the Abelian case that the action of the global gauge group can literally be viewed as the free and transitive right action of $G$ on $P$.\endnote{A gauge transformation is a bundle automorphism, but in the non-Abelian case the free and transitive group action of $G$ on $P$ need not actually yield bundle automorphisms. To see this, take $h_1,h_2\in G$ that do not commute and consider the map $f\colon P\to P$ given by $p\mapsto p h_1$. Then we have $f(p h_2)=(p h_2)h_1=p(h_2h_1)\neq p(h_1h_2)=f(p)h_2$.}

(2)~~{\it Interpretative aspects of gauge symmetries. }We will now clarify, in general terms, why global gauge transformations are physical. For they are a subgroup of the total gauge group $\mathcal{G}$, and so one might naively think that, if the transformations in $\mathcal{G}$ are unphysical, mathematical redundancies, then so are the global gauge transformations (and this point is close to the view (ii) of the Higgs mechanism that we discussed in Section \ref{intro}). However, it not true that the whole gauge group $\mathcal{G}$ is unphysical. For it has been shown that, in the presence of boundaries, the subgroup of global gauge transformations {\it is} physical. This is the upshot of the constrained Hamiltonian analysis, which provides a formal correlate of the meaning of `gauge' in the sense of `redundant', which constrasts with the `formal' characterisation of `gauge' as follows (see \autocite{tehGalileoGaugeUnderstanding2016}):

\textit{Redundant}: A gauge transformation of the fields that has no physical effect, i.e.~it does not lead to distinct empirical predictions.

\textit{Formal}: A gauge transformation as defined in mathematical terms, i.e.~an element of the gauge group $\mathcal{G}\cong C^\infty(M,G)$ (where $M$ is spacetime and $G$ is a compact matrix Lie group) that acts on the fields. 

As \autocite{tehGalileoGaugeUnderstanding2016} emphasizes, confusion ensues from mixing these two connotations of the word `gauge'. The statement that a gauge transformation is \textit{redundant} is an interpretative statement, while \textit{formal} is a mathematical i.e.~non-interpretative statement. In this paper, we always use the word `gauge' in the sense of \textit{formal}, i.e.~in the sense of an automorphism of a principal bundle $P$. Furthermore, we will contrast the class of \textit{redundant} gauge transformations with the class of gauge transformations that have: 

\textit{Direct empirical significance (DES):} A gauge transformation of the fields has DES if it has a physical effect, i.e.~it leads to distinct empirical predictions.

\subsection{Empirical significance of global gauge symmetries}\label{sec:DESglobal}

In the recent philosophical literature, there has been substantial debate about whether, and if so how, gauge symmetries can exhibit DES in so-called Galileo's ship scenarios.\endnote{See \autocite{kossoEmpiricalStatusSymmetries2000,bradingAreGaugeSymmetry2004,healeyPerfectSymmetries2009,Rovelli:2013fga,greavesEmpiricalConsequencesSymmetries2014,friederichSymmetryEmpiricalEquivalence2015,ramirezAbandoningGalileoShip2021,wallaceIsolatedSystemsTheireen,wallaceIsolatedSystemstwee}.} It is important to note that DES in such a scenario is \textit{relational}, i.e. it refers to empirical significance when a subsystem undergoes a symmetry transformation in relation to an untransformed environment. It has been shown in various ways that the global gauge group can indeed exhibit DES in this sense. For instance, the \textit{'t Hooft beam splitter} is a thought experiment that showcases the physicality of global gauge transformations in electromagnetism when they act on a scalar field assumed to vanish on the boundary of a subsystem \autocite[]{greavesEmpiricalConsequencesSymmetries2014}. In addition, a number of authors have used sophisticated mathematical methods to derive the physicality of global gauge transformations.\endnote{For a sample of the most recent literature, see \autocite{gomesUnifiedGeometricFramework2019,gomesQuasilocalDegreesFreedom2021,gomesHolismEmpiricalSignificance2021,gomesWhyGaugeConceptual2022,gomesGaugeTheoryGeometrisation2025}. For a more detailed analysis of these articles in relation to our ideas on the Higgs mechanism we refer the reader to Section 4.3.4 of \autocite{borsboom2024spontaneous}.} A notable result\endnote{However, this is a result on subsystems with a boundary in between, not on \textit{asymptotic} boundary conditions.} that supports our analysis is Theorem 1 from \autocite{gomesHolismEmpiricalSignificance2021}, which can be rephrased as:

\begin{theorem}[Rigid variety for $U(1)$]\label{gomestheorem}
    For the Maxwell theory as coupled to a Klein-Gordon scalar field in a simply-connected universe: given the physical content of two regions, for matter vanishing at the boundary but not in the bulk of the regions, the universal state is undetermined, resulting in a residual variety parametrized by an element of $U(1)$. Here the particular action of $U(1)$ is that which leaves the gauge fields, but not the matter fields, invariant.
\end{theorem}

The results discussed in this literature thus suggest that DES is indeed exhibited by the {\it global gauge transformations}, which transform a subsystem and act in particular on the boundary.\endnote{Hence the relation to edge modes, see e.g. \autocite{donnellyLocalSubsystemsGauge2016,carrozzaEdgeModesReference2022,rielloEdgeModesEdge2021}} 
The analogy is here with Galileo's ship, where a transformation that is made locally on the system (i.e.~the ship) does not change the physics inside the system, even though it does change the physical state with respect to an environment outside that boundary. The difficulty with the Higgs mechanism, however, is that it is usually applied to the universe as a whole, and one does not imagine the universe to be a subsystem of anything. The boundaries of the universe are not finite boundaries delimiting the passage into an environment, but asymptotic or conformal boundaries. The environment is located ``at infinity". This means that boundary gauge symmetries in this context are \textit{asymptotic symmetries}, which have been intensively studied both for gravity as well as Yang-Mills theory.\endnote{See \autocite{bondiGravitationalWavesGeneral1962,sachsAsymptoticSymmetriesGravitational1962,stromingerAsymptoticSymmetriesYangMills2014,henneauxAsymptoticSymmetriesElectromagnetism2018,henneauxAsymptoticStructureGravity2020}. For a philosophical discussion of this topic in general relativity, see \autocite{haroInvisibilityDiffeomorphisms2017}.} In \autocite{borsboom} the global gauge group is shown to be the asymptotic symmetry group of Yang-Mills theory on a Cauchy surface.

One might therefore worry that there is a contradiction between claiming that global gauge symmetries, viewed as an asymptotic symmetry group, can exhibit DES, while at the same time accepting the central claim of \autocite{greavesEmpiricalConsequencesSymmetries2014} that symmetries of the whole universe have no empirical significance. The logic seems the same global gauge symmetries as for other symmetries: translations of the whole universe are also taken to be unphysical (at least from a Leibnizian viewpoint), though they are the archetypal example of a symmetry with DES, but understood relationally across subsystems. Global gauge symmetries can indeed be applied to subsystems, in which case we \textit{know} them to be physical, as shown empirically by the Josephson current in superconductors. This leads physicists to consider \textit{domain formation} in spontaneously broken gauge theories, which occurs when the Higgs field breaks global gauge symmetry differently in different regions of the universe. One then expects to find defects along the boundaries of these domains, known as \textit{cosmic strings} \autocite{Kibble:1976sj,Vilenkin:2000jqa}. Thus global gauge symmetries clearly exhibit DES when applied to subsystems, and this is taken seriously as an empirical possibility by prominent physicsists. This in itself is enough to support the claim that global gauge symmetry breaking is the physical content in the Higgs mechanism. 

Nonetheless, it would still seem that global gauge symmetries can have no physical significance when applied to the universe \textit{as a whole}, as is the case in the Abelian Higgs model on Minkowski spacetime. We, however, resist this claim for two reasons. We espouse both, but our own position aligns with the second.

Firstly, one can simply understand the asymptotic fall-off conditions on the fields in the Abelian Higgs model, which are necessary for a well-defined Lagrangian and lead to the asymptotic global gauge group, as an \textit{idealization} of a perfectly isolated subsystem. Indeed, on the scale of an atom, a few meters away seems like spatial infinity, and those few meters can be modelled by Minkowski spacetime, including the necessary fall-off conditions. For a more general analysis of the role of idealized fall-off conditions in symmetry breaking, see \autocite{borsboomdupont}.

Secondly, we believe that the very assumption in \autocite{greavesEmpiricalConsequencesSymmetries2014} of the unphysicality of symmetries applied to the whole universe is false when an asymptotic symmetry group is present, such as the global gauge group in Yang-Mills theory on Minkowski spacetime. This is because, in the presence of asymptotic boundary conditions, the ``whole universe'' is effectively divided into a ``bulk'' and a ``boundary''. Indeed, much of current theoretical high-energy physics revolves around asymptotic gauge symmetries, especially in (celestial) holography. Physicists do not take such asymptotic symmetries to be unphysical, even though they know very well that these act on a boundary ``at infinity'', i.e. a conformal boundary. In part, this is because asymptotic symmetries do not lead to a breakdown of determinism as do bulk symmetries. This fact can be seen by carefully studying the symplectic structure of a gauge theory in the presence of boundary conditions, which is the focus of Section \ref{electrogauss}. But at a deeper level, it is mistaken to think that an asymptotic symmetry group act only ``at infinity''. Asymptotic boundaries encode fall-off behavior, made precise by means of conformal compactifications \autocite{borsboom}. The statement that an asymptotic symmetry group ``acts at infinity'' really means that \textit{it acts on the part of the field that can be detected through any boundary by means of Gauss' law.} In electromagnetism, this component is known as the \textit{electric flux}, and it is the best-known example of a so-called \textit{surface charge}. The electric flux equals the distribution of charges in space by Gauss' law. In particular, this holds for the flux ``at infinity'', i.e. through the conformal boundary. The global gauge group, understood as the asymptotic symmetry group of electromagnetism, is the generator of this electric flux surface charge. One must not think of the action of that group as occurring only ``at infinity`` and therefore as unphysical, but rather as acting everywhere in space on those components of the electric field that fall off sufficiently slowly to result in a detectable flux arbitrarily far away. This is a form of non-locality, which we will see in Section \ref{higgsqft} to imply non-locality in the sense of a failure of microcausality when the Abelian Higgs model is quantized. In that Section we will also see that the global gauge group is responsible for the superselection of electric charge in the quantized theory, which is also an empirically detectable phenomenon.

Before we move on we recall that the philosophical literature has also considered the indirect empirical significance of a gauge transformation:\endnote{See e.g.~\autocite[p.~648]{bradingAreGaugeSymmetry2004}}

\textit{Indirect empirical significance (IES)}: A gauge transformation of the fields has IES if it does not itself have a physical effect, i.e.~the different states that it relates do not themselves lead to distinct empirical predictions, but the properties of the fields and laws that are connected with the existence of this symmetry do have DES. In the relational understanding of the physical significance of symmetries, any symmetry transformation applied to a whole system without boundaries can only have IES, so the non-trivial case is when even subsystem transformations do not lead to distinct empirical predictions as viewed from the environment.

In other words, the physical significance of this gauge transformation is indirect, through its connection with fields and laws that do have DES. Although the symmetry transformation itself is not observable, its existence does have observable consequences. A well-known example is the local gauge symmetry of the Maxwell theory: although (as we will discuss) it has no DES, it does have IES in that it implies that the Maxwell theory can only be coupled to a current that is conserved. Thus it implies the physically significant fact of charge conservation, which is tested and used in e.g. accelerator experiments. Therefore, gauge symmetries that are \textit{redundant} can, and often do, have IES. When electromagnetism is formulated entirely in terms of gauge-invariant observables like electric and magnetic fields, then it is unclear why currents of matter fields should be conserved, as coupling in the Lagrangian or Hamiltonian formalism can only be achieved through the introduction of the gauge potential. This is a version of the traditional `gauge argument' \autocite{gomesGaugeTheoryGeometrisation2025}.

\subsection{The quotient group of physical symmetries}\label{gps}

Sections \ref{gfgt} and \ref{sec:DESglobal} argued, on general grounds, that the gauge transformations with DES are the global ones. The aim of this Section is to explain how the global (i.e. rigid) gauge transformations arise \textit{precisely} as the physical gauge group by taking the quotient of all \textit{boundary-preserving} symmetries by all \textit{redundant} symmetries.

Descriptions of physical systems usually come with boundary conditions. If the Lagrangian or Hamiltonian of a system exhibits a symmetry, then it is important that the symmetry group acts on that system in such a way that the boundary conditions are preserved to guarantee that the description of the system is well-defined. In particular, if we consider gauge theories with boundary conditions, then we must ensure that the action of the gauge group is boundary-preserving, i.e. respects the boundary conditions on the fields in the theory. In what follows, we denote the group of boundary-preserving gauge transformations by $\mathcal{G}^I$, where the notation $I$ stands for `invariant',  since these are the transformations that leave the boundary conditions invariant.

Thus, the group of physical symmetries equals all those elements of $\mathcal{G}^I$ that are not \textit{redundant}. But how can we characterize precisely which gauge transformations are redundant and which ones are not? To this end, one can make use of the \textit{constrained Hamiltonian formalism},\endnote{The go-to physics text is \autocite{henneauxQuantizationGaugeSystems1992}. A much more mathematically, though not so well-known, book on this topic is \autocite{binz}, which we use frequently. The state of the art is presented in \autocite{gotayMomentumMapsClassical2004,gotayMomentumMapsClassical2004a,Gotay2006MomentumMA}.} which originated in \autocite{diracGeneralizedHamiltonianDynamics1950,PhysRev.83.1018}. This formalism allows one to straightfowardly identify \textit{redundant} gauge transformations by investigating which transformations are generated by the constraints of the theory. Such transformations are directly seen to lead to indeterminism, in that the orbits of these transformations are null directions of the symplectic form, giving multiple possible solutions of Hamilton's equations. The generation of gauge transformations by the Hamiltonian constraints occurs by taking Poisson brackets (see Section \ref{electrogauss}). The identification of redundancy, in the sense of a breakdown of determinism, with generation by constraints can also be viewed in the light of Noether's \textit{second} theorem: the time-localizability of gauge transformations simultaneously leads to indeterminism by the second Noether theorem \autocite{Wallace_2003}, as well as to the Hamiltonian constraint that needs to be imposed to remove this indeterminism. 

This, then, we take to be the fundamental notion of `gauge': time-localizable symmetry transformations that lead, through Noether's second theorem, to Hamiltonian constraints on the fields. For Yang-Mills theory the relevant constraint is the so-called \textit{Gauss law constraint}, which is just the usual Gauss law $\nabla\cdot \textbf{E}=0$ in the special case of electromagnetism. We will explain precisely how constraints arise and generate gauge transformations in Section \ref{electrogauss}. The remainder of this Section is dedicated to the introduction of various subgroups of the formal gauge group $\mathcal{G}$ which will appear frequently throughout the article.
\\
\\
{\it Gauge transformations that are generated by the Gauss constraint.} To identify the gauge transformations that are generated by the Gauss law constraint, \autocite{tehGalileoGaugeUnderstanding2016} quotes a result in  \autocite{balachandranGaugeSymmetriesTopology1994} that states that the asymptotic condition that secures that a gauge transformation $g\in\mathcal{G}$ can be generated by the Gauss constraint is that it approaches unity at infinity at the appropriate rate:
\begin{align}\label{gaugetranscondition}
    g(\textbf{x})\to 1\;\;\;\;\;\text{as}\;\;\;\;\;|\textbf{x}|\to\infty\,,
\end{align}
where ${\bf x}\in\mathbb{R}^3$. However, this statement was rather imprecise and has been rederived completely rigorously, including precise fall-off rates, in \autocite{borsboom}. We denote the subgroup of maps $M\to G$ that asymptote to the identity by $\mathcal{G}^\infty$, a notation used already in \autocite{Friedman:1983at,Friedman:1983au}. It is important to note that, in order to even be able to speak of the identity at infinity, we need a section or frame, i.e.~a trivialization, of the bundle at infinity. In fact, condition \eqref{gaugetranscondition} already assumes a trivialization of the the bundle everywhere, since a gauge transformation can only be viewed as a map $\Sigma\to G$ when one has a section $\Sigma\to P$. One may worry, therefore, that imposing this condition already amounts to some form of breaking of gauge-invariance. However, we must keep in mind that a choice of section actually induces an \textit{isomorphism} between $\mathcal{G}(P):=\text{Aut}(P)$ and $C^\infty(\Sigma,G)$. Thus, condition \eqref{gaugetranscondition} induces a gauge-independent definition of the subgroup $\mathcal{G}^\infty$ by translating this condition back to $\mathcal{G}(P)$ through the isomorphism with $C^\infty(\Sigma,G)$. The form of the isomorphism depends on the choice of section, but the resulting subgroup $\mathcal{G}^\infty\subset \mathcal{G}(P)$ does not. Of course, this does assume that $P$ is trivializable, but on a Cauchy surface isomorphic to $\R^3$ that is automatically the case.

In addition, the transformations generated by the Gauss constraint are \textit{small}, i.e.~homotopic to the identity map, $x\mapsto 1\in\mathcal{G}$. This is so because the Gauss constraint smears elements of the infinite-dimensional Lie algebra $\text{Lie}(\mathcal{G})$ of the gauge group $\mathcal{G}$, and these must be exponentiated to get an element of the Lie group $\mathcal{G}$ itself. And it is a well-known fact that the image of the exponential map lies in the connected component of the identity of the Lie group, which explains why the Gauss constraint generates only small gauge transformations. Indeed, configurations connected by large gauge transformations are empirically distinguishable by e.g. instanton numbers. For another example of the empirical significance of \textit{large} gauge transformations see \autocite{gomesEliminativismQCDtheta2024}.

We denote by $\mathcal{G}^\infty_0\subset\mathcal{G}$ the subgroup of gauge transformations $g$ that satisfy the above two conditions, i.e.~they: \\

(i)~~are small, i.e. $g\in \mathcal{G}_0$,

(ii')~~satisfy the asymptotic condition Eq.~\eqref{gaugetranscondition}. \\
\\
Since these are generated by the Gauss constraint and therefore are responsible for the indeterminism typical of time-localizable symmetries, the gauge transformations $\mathcal{G}^\infty_0$ are the {\it redundant} ones, i.e.~the ones that are not physical. (Anticipating that we will replace the condition (ii') by the more general (ii) below, we gave this label a tilde.)

Smallness, i.e.~condition (i), is automatically satisfied for the U(1) structure group of electromagnetism in three spatial dimensions, but it is not automatically satisfied for the non-Abelian structure group $SU(2)$: and so, in general, (i) and (ii') are independent conditions.\\
\\
{\it Gauge transformations that are not generated by the Gauss constraint.} To find the set of all admissible gauge transformations, i.e.~not only those that are redundant but also those that have DES, we should drop the smallness condition (i). Also, we should replace (ii') by a more general asymptotic condition that is not specific to those gauge transformations that are generated by constraints, but which only makes sure that gauge transformations are boundary-preserving. Thus we consider the following gauge transformations, which, as proved in \autocite{borsboom}, leave invariant the asymptotic boundary conditions of fields on a Cauchy slice of Minkowski spacetime:

(ii)~~the subgroup of gauge transformations that are constant at infinity, i.e.~$\exists g_0\in G:~~~g({\bf x})\rightarrow g_0$ as $|{\bf x}|\rightarrow\infty$.

Clearly, (ii') is a special case of (ii) with $g_0=1$. We denote by $\mathcal{G}^I$ the subgroup of transformations that satisfy condition (ii), but not necessarily condition (i). Because these gauge transformations preserve the boundary conditions, we say that they are `boundary-preserving'.

The reason we need to impose condition (ii) on gauge transformations is that boundary conditions are an integral part of how we represent a physical system: they restrict the types of configurations that are allowed. A transformation that does not respect the boundary conditions can therefore not be viewed as a symmetry of a physical system (in either the senses of {\it redundant}, or of DES that are of our interest here), since it maps the fields describing one physical system into those of a different physical system, and in some cases even into a set of fields that do not correspond to a physical system at all (e.g.~because they are inconsistent with the theory's dynamics). Specifically, in field theory the boundary conditions are chosen such that physical quantities, in particular the total energy and action, are finite, and we will assume that the conditions (ii) are of this type. For mathematical details on the precise implementation of such boundary conditions we refer to \autocite{borsboom,borsboominstantaneous}.

To sum up: $\mathcal{G}^\infty_0$ is the subgroup of gauge transformations that satisfy (i) and (ii') (and therefore also (i) and (ii)), $\mathcal{G}^\infty$ satisfy (ii') (and therefore also (ii)), and $\mathcal{G}^I$ satisfy (ii). Thus we have the following hierarchy of subgroups \autocite{tehGalileoGaugeUnderstanding2016}:
\begin{align*}
    \mathcal{G}^\infty_0\subset\mathcal{G}^\infty\subset\mathcal{G}^I\subset\mathcal{G}\,.
\end{align*}

By the analysis of \autocite{tehGalileoGaugeUnderstanding2016}, which replaces the condition of \textit{non-interiority} from \autocite{greavesEmpiricalConsequencesSymmetries2014} by non-redundancy, the symmetries that exhibit DES are those that are boundary-preserving but non-redundant, i.e.~not generated by the Gauss law constraint. Thus we obtain:
\begin{align}\label{GDES}
\mathcal{G}_\text{DES}=\mathcal{G}^I/\mathcal{G}^\infty_0\,.
\end{align}

However, the derivation showing that, for Yang-Mills theory on a Euclidean Cauchy slice, this group indeed equals the global gauge symmetry group (or a copy thereof for every homotopy class of gauge transformations), was only recently made rigorous\autocite{borsboom}. The main point made in that article is that the requirement that boundary-preserving gauge transformations $g\in\mathcal{G}^I$ become constant at asymptotic infinity follows not only from the boundary conditions on the fields, but also from the structure of the instantaneous state space of classical field theories in the Lagrangian formulation. Studying the instantaneous state space of Yang-Mills-Higgs theory, it can be shown that $\mathcal{G}^I$ is different in the unbroken and broken phases. In the unbroken phase $\mathcal{G}^I$ consists of all asymptotically constant gauge symmetries, like for pure Yang-Mills theory. But in the broken phase, i.e. when we take a minimum of the Higgs potential $V(\varphi)$ to have zero energy, rather than $\varphi=0$, the boundary-preserving gauge symmetries $\mathcal{G}^I$ are all transformations that become the identity at infinity. Thus, we find that $\mathcal{G}_\text{DES}$ consists of a copy of the global gauge group for every homotopy class only in the unbroken phase. It is trivial/discrete in the broken phase.

We will use these results throughout this article, while referring to \autocite{borsboom,borsboominstantaneous} for the mathematical details on how to take the quotient in Eq.~\eqref{GDES}. However, we do need to understand how the group of redundant gauge transformations $\mathcal{G}^\infty_0$ arises by studying which transformations are generated by the Gauss law constraint. The next Section will therefore be dedicated to this. Doing so will provide us with the necessary tools to give our harmonized account of gauge symmetry breaking in the Higgs mechanism in Section \ref{coulomb}.

\section{The Gauss Law Constraint in Electromagnetism}\label{electrogauss}

Now that we understand what the group $\mathcal{G}^I$ looks like in pure electromagnetism (i.e. satisfying Eq.~\eqref{gaugetranscondition}) and the unbroken and broken phases of the Abelian Higgs model, we turn to the group $\mathcal{G}^\infty_0$ of redundant gauge transformations. The aim of this Section is to show that this group consists of gauge transformations that become the identity at asymptotic infinity. We first explain how gauge transformations relate to constraints in Section \ref{ggsc}. We then specialize to electromagnetism in Section \ref{electromag}. Finally we show how the Coulomb gauge arises as the natural gauge for studying global gauge symmetry breaking in Section \ref{coulombgauge}.

\subsection{Redundant gauge symmetries are generated by constraints}\label{ggsc}

{\it Constraints in the Hamiltonian formalism generate redundant gauge transformations because we require determinism.} While constraints can arise in the Hamiltonian formalism for a number of reasons, for us, since we are interested in gauge theories, the appearance of constraints will signal the presence of a redundant symmetry: in other words, it will signal the failure of the configuration space to be in a one-to-one correspondence with the space of physical states. And, as we discuss below, we will define the latter in terms of determinism under time evolution. Thus, empirical significance and determinism are linked: assuming the system's physical state to evolve uniquely and deterministically, we must conclude that a gauge transformation is unphysical if it leads to a breakdown of determinism, but if it does not we are \textit{not entitled} to conclude this about said transformation. Since we will see that global gauge transformations in electromagnetism are \textit{not} linked to indeterminism when appropriate asymptotic fall-off conditions are imposed, we should not take them to be unphysical.

In the Lagrangian formalism, if the Hessian matrix of the Lagrangian has a zero eigenvalue, so that its determinant is zero,\endnote{The Hessian matrix of a Lagrangian $L$ is the matrix $\partial^2L/\partial\dot q^i\partial\dot q^j$. For details, see \autocite[pp.~4-5]{henneauxQuantizationGaugeSystems1992}.}
the acceleration is not uniquely determined by the positions and velocities, and the solutions contain arbitrary functions of time \autocite{henneauxQuantizationGaugeSystems1992}. This implies that, in such a situation, the time evolution is in general not deterministic. In a gauge theory, this lack of determinism, i.e.~the vanishing of the determinant of the Hessian, is the consequence of a time-localizable gauge symmetry \autocite{gotayMomentumMapsClassical2004a,Wallace_2003}.

From this perspective, the constrained Hamiltonian formalism is a way to construct a deterministic theory with a unique time evolution, by identifying points on the phase space that are related by a redundant symmetry. The submanifold of points on phase space that are related by a redundant symmetry, and that thus correspond to the same physical state, is generated by a set of constraints. Hence the slogan `constraints generate redundant gauge transformations', which we will take as our criterion for {\it redundant} gauge transformations. 

Note that this criterion has, as it of course should, a crucial interpretative aspect. This is because our consideration of constraints as generating redundancies is directly motivated by the requirement of having a {\it deterministic theory with a unique time evolution}. This thus corresponds to the second notion of `gauge' in \autocite{struyveGaugeInvariantAccounts2011a}. Indeed, if one drops this latter requirement, then the gauge transformations that are generated by constraints are {\it not} in general redundant. Alternatively, if one uses a different dynamical principle to define one's set of physical states, one will in general have a different criterion for a gauge transformation being {\it redundant} (see \autocite{henneauxQuantizationGaugeSystems1992},
\autocite[p.~149]{earmanTrackingGaugeOde2003}).\endnote{For an example of a treatment of a gauge theory where one gives up the requirement of determinism, see \autocite[pp.~88-90]{belot1995determinism}.}\\
\\
{\it The constraints give the set of points in the cotangent bundle where the Lagrangian transformation is invertible.} Mathematically speaking, constraints arise because the Legendre transformation is not a diffeomorphism. More specifically, constraints relate to the fact the velocities in the Lagrangian formalism cannot be obtained from the canonical momenta in the Hamiltonian formalism. In such a case, the canonical momenta are not independent, but satisfy constraint relations. We now fill in some of the mathematical details.

Recall that, in the Lagrangian formalism, we work with positions and velocities, i.e.~pairs $(q^i,\dot q_i)\in TQ$ (in field theories these are field configurations and their time derivatives), which take values in the tangent bundle, $TQ$, to the configuration space, with $q^i\in Q,\dot{q}^i\in T_qQ$. By contrast, in the Hamiltonian formalism, we work with positions and momenta, i.e.~pairs $(q^i,p_i)\in T^*Q$, which take values in the cotangent bundle, $T^*Q$.

Thus the Legendre transformation is a map, $TQ\to T^*Q$, from the tangent bundle to the cotangent bundle. In simple cases, this map is invertible, i.e.~{\it hyperregular}, so that it is in fact a diffeomorphism of the tangent bundle. However, in cases where the Lagrangian has e.g.~internal symmetries, this map need not be hyperregular \autocite[p.~186]{marsdenIntroductionMechanicsSymmetry1999}. 

If the Legendre transformation is not surjective, then there are relations $c_m(q^i,p_j)=0$ between the momenta $p_j$ and generalized positions $q^i$, called \textit{constraints}. A constraint expresses an equivalence relation of a set of points in the cotangent bundle that, under the Legendre transformation, do not have a unique inverse in the tangent bundle. Since these are precisely the points where determinism fails due to a gauge symmetry, by interpreting them as corresponding to the same physical state, we restore determinism. As a consequence, the corresponding gauge transformation is {\it redundant}. Thus, in the Hamiltonian formulation, we may wish to quotient our constraint surface by the {\it redundant} gauge transformations.\endnote{A precise treatment of this reduction procedure, even in the presence of boundaries, is given in e.g. \autocite{binz,rielloHamiltonianGaugeTheory2024,rielloNullHamiltonianYang2025}}. The best known example of a gauge constraint is undoubtedly Gauss's law from electromagnetism, which we will discuss in Section \ref{electromag}.

The constraint relations define the \textit{primary constraint set} $C\subset T^*Q$, which is the image of the Legendre transformation (provided the Lagrangian $\mathcal{L}$ is \textit{regular} \autocite{marsdenIntroductionMechanicsSymmetry1999}). Primary constraints do not rely on the equations of motion, while secondary constraints arise as the requirement that the primary constraints be preserved in time as the system evolves. This can be iterated to obtain tertiary constraints etc. \autocite{healeyGaugingWhatReal2007}. The space that is defined by requiring all constraints to be satisfied is called the \textit{constraint surface}, and it is assumed to be a submanifold smoothly embedded in phase space.\endnote{In Yang-Mills theory, the constraint surface is given by $\mathcal{C}=J_\mathfrak{h}^{-1}(0)$, where $J_\mathfrak{h}$ is the momentum map of the group of infinitesimal localizable gauge symmetries, whose Lie algebra is $\mathfrak{h}$ \autocite{binz}.}

If a function $F$ on phase space vanishes on the constraint surface, it is said to vanish \textit{weakly}, written $F\approx 0$. Constraints whose Poisson brackets with all other constraints vanish weakly are called \textit{first class}. Dirac's insight was that ``first class constraints generate gauge transformations'' (where `generating' means taking Poisson brackets), although there are various subtleties concerning this statement that we do not consider here, see e.g. \autocite{henneauxQuantizationGaugeSystems1992,earmanTrackingGaugeOde2003,Gotay2006MomentumMA}.\endnote{Pitts has objected to this slogan \autocite{pittsFirstClassConstraint2014}, but this objection has again been objected to \autocite{pooleyFirstclassConstraintsGenerate2022}.} These subtleties are not relevant to our investigation, since they do not play a role in Yang-Mills theory in temporal gauge, where the Gauss law is the only constraint. The Gauss law in Yang-Mills theory is known to generate precisely all local gauge transformations \autocite{binz,Gotay2006MomentumMA}.

To better understand whence comes this notion of \textit{gauge} as `generated by the first class constraints', it is useful to consider the symplectic formalism underlying constrained Hamiltonian analysis.\endnote{For a conceptual overview see \autocite{gomesHowChooseGauge2024}.} The phase space cotangent bundle $T^*Q$ is a symplectic manifold, i.e.~a manifold equipped with a closed non-degenerate 2-form $\omega\in\Omega^2(T^*Q)$. It is written locally as
\begin{align}\label{omegac}
    \omega=\sum_i dq^i\wedge dp_i\,.
\end{align}
On a symplectic manifold $(\mathcal{M},\omega)$, every smooth function $H\in C^\infty(\mathcal{M})$ defines a Hamiltonian vector field $X_H\in\mathfrak{X}(\mathcal{M})$, where $\mathfrak{X}(\mathcal{M})$ denotes the set (more precisely: the $C^\infty$-module) of vector fields on $\mathcal{M}$, through
\begin{align*}
    dH=\omega(X_H,\cdot)\,.
\end{align*}
Here the $\cdot$ denotes the vacant spot in which an arbitrary vector field from $\mathfrak{X}(\mathcal{M})$ can be inserted. Note that $dH$ also eats such a vector field by definition of the differential operator $d:C^\infty(\mathcal{M})\to \Omega^1(\mathcal{M})$ which sends a smooth function to a 1-form. The Poisson bracket between $f,g\in C^\infty(\mathcal{M})$ is then defined as 
\begin{align}\label{omegaX}
    \{f,g\}=\omega(X_f,X_g)\,.
\end{align}
It is not difficult to see that for the canonical 2-form on the cotangent bundle this locally gives the familiar Poisson bracket:
\begin{align}\label{Poisson}
    \{f,g\}=\sum_i\frac{\partial f}{\partial q^i}\frac{\partial g}{\partial p_i}-\frac{\partial f}{\partial p_i}\frac{\partial g}{\partial q^i}\,.
\end{align}
The idea behind Hamiltonian vector fields is that their integral curves are precisely the trajectories through phase space satisfying Hamilton's equations. Similarly, we can consider the Hamiltonian vector fields $X_{c_m}$ associated with the constraints $c_m\colon \mathcal{M}\to\R$. If all constraints are first class, then these vector fields flow tangentially to the constraint surface. If not all constraints are first class, then one can use the Dirac algorithm to guarantee so. We will not be concerned with that possibility since in our case of interest all constraints are first class.

If all constraints are first class then their associated vector fields are null directions of the symplectic form \textit{on the constraint surface}, i.e.~$\omega(X_{c_m},X_{c_n})=\{c_m,c_n\}\approx 0$. The existence of such null directions is possible because $\omega$ can be degenerate on the constraint surface, whereas it is by definition non-degenerate on the entire phase space. It is for this reason that one calls these directions \textit{gauge} and their integral curves \textit{gauge orbits}. Points within one orbit are interpreted as physically equivalent, so any physical quantity must be gauge-invariant in the sense of being constant on every gauge orbit. 

In addition to calculating the Poisson bracket, Eq.~\eqref{Poisson}, between the constraints themselves, we can also use it to calculate the Poisson bracket of a constraint with a field $F$. Schematically, we write:
\begin{align}\label{Fcm}
\{F,c_m\}=\delta_m F\,,
\end{align}
where $F$ is any field of the theory, and $c_m$ is any first-class constraint. The result is then that $\delta_m F$ is the corresponding {\it redundant} gauge transformation of $F$ \autocite[p.~17]{henneauxQuantizationGaugeSystems1992}. Thus if $F$ is a gauge-invariant field, the right-hand side of this equation is zero. And if $F$ is a non-gauge-invariant field, the right-hand side is its infinitesimal gauge variation. Thus by taking Poisson brackets with constraints, we move along the orbit of a {\it redundant} gauge symmetry in the phase space. And by exponentiating such infinitesimal variations, we can generate finite gauge orbits. In the next Section,  we will illustrate this in the example of the Maxwell theory.

\subsection{Electromagnetism}\label{electromag}

In this and the next Section we show explicitly how the Gauss law constraint generates redundant gauge transformations in electromagnetism (i.e.~$G=U(1)$), and how this leads to the Coulomb gauge as the appropriate gauge to handle the physical gauge group derived in the previous Section.\endnote{In this respect we disagree with \autocite{wallaceGaugeInvarianceGauge2024}, where it is argued that unitary gauge is appropriate for electrodynamics with a Higgs field.} We will need these results and explicit expressions in Section \ref{coulomb} for our harmonization of the two conflicting gauge-invariant accounts of the Higgs mechanism anticipated in Section \ref{intro}. The Coulomb gauge precisely removes the redundant gauge transformations generated by the Gauss law constraint, but it leaves the physical gauge group of global gauge symmetries. We will therefore use it to study global gauge symmetry breaking in the Abelian Higgs mechanism, combining ideas from \autocite{struyveGaugeInvariantAccounts2011a} and the DFM \autocite{berghoferGaugeSymmetriesSymmetry2023}.

From now on we work on a trivial bundle $M\times U(1)$ over spacetime.  The first thing to do is to find the constraints of electromagnetism. The covariant Lagrangian of pure Maxwell theory is:
\begin{align*}
    \mathcal{L}=-\frac{1}{4}F^{\mu\nu}F_{\mu\nu}\,,
\end{align*}
to which the Lagrangian for a charged scalar field can be added. Here and from now on we use the Einstein summation convention for pairs of indices of which one is up and one is down. The Legendre transformation gives the following momenta canonical to the gauge field:
\begin{align}\label{pii}
    \Pi^i&=\frac{\partial\mathcal{L}}{\partial \dot{A}_i}=-\frac{1}{2}F^{\mu\nu}\frac{\partial}{\partial \dot{A}_i}\left(\partial_\mu A_\nu-\partial_\nu A_\mu\right)=-\frac{1}{2}F^{\mu\nu}(\delta^0_\mu\delta^i_\nu-\delta^0_\nu\delta^i_\mu)
    \\
    &=-\frac{1}{2}(F^{0i}-F^{i0})=F^{i0}=-E^i\,.
\end{align}
Clearly $\Pi^0=0$ because $F^{00}=0$. This indicates that $A_0$ is a Lagrange multiplier. Thus $\Pi^0=0$ is the primary constraint. If one then imposes $\{\Pi^0,H\}=0$, one finds the Gauss law, i.e.~$\rho=J^0=\nabla\cdot \textbf{E}$, where $\rho$ is the charge density and $\textbf{E}$ the electric field. This is the secondary constraint.

For ease of presentation we set $A_0=0$, a procedure know as \textit{temporal gauge}. Combining this with a spacetime split $\Sigma\times\R$, where $\Sigma\cong \R^3$ is a Cauchy, the gauge fields become functions $A_i\in C^\infty_I(\Sigma )$ on space. These must approach zero sufficiently quickly, so we again use the notation I to indicate behavior relating to asymptotic boundary conditions. In this case it means that we consider fields which vanish at infinity, whereas before it signified gauge transformations that respect asymptotic boundary-conditions. Similarly $E^i\in C^\infty_I(\Sigma )$, and phase space consists of such pairs $(\textbf{A},\textbf{E})$. We can then use a smooth function $\lambda\colon \Sigma \to\R$, again understood to be a Lagrange multiplier, to obtain the \textit{smeared Gauss constraint}
\begin{align}\label{Gaussc}
    G_\lambda =\int_\Sigma  d^3x\,\lambda\,(\nabla\cdot\textbf{E}-\rho)\,,
\end{align}
which is a function on the phase space. Here the gauge transformation parameter $\lambda$ is an element of the infinite-dimensional Lie algebra $\text{Lie}(\mathcal{G}^I)$ of the boundary-preserving gauge group. 

We are interested in calculating how the smeared Gauss constraint acts on the fields, i.e.~in its Poisson brackets with $A_i$ and $E^i$. The symplectic form is the canonical one:
\begin{align}\label{symplecticformmaxwell}
    \Omega=\int_\Sigma  d^3x~\mathbb{d}A_i\wedge \mathbb{d}E^i\,,
\end{align}
where we follow \autocite{gomesQuasilocalDegreesFreedom2021,gomesHowChooseGauge2024} in using the double struck $\mathbb{d}$ to indicate that this is the differential operator on the infinite-dimensional phase space coordinatized by the fields $A_i$ and $E^i$, and not on the three-dimensional space $\Sigma $, for which we use the usual $d$. From the expression for the symplectic form $\Omega$, it follows that the Poisson bracket of two functionals $F$ and $G$ of the fields is given by
\begin{align}\label{Poissonb}
    \{F,G\}=\int_\Sigma  d^3x\left(\frac{\delta F}{\delta A_i(\textbf{x})}\frac{\delta G}{\delta E^i(\textbf{x})}-\frac{\delta F}{\delta E^i(\textbf{x})}\frac{\delta G}{\delta A_i(\textbf{x})}\right),
\end{align}
where $\delta/\delta A_i(\textbf{x})$ and $\delta/\delta E^i(\textbf{x})$ denote the functional derivatives of phase space functions with respect to the fields $A_i(\textbf{x}), E^i(\textbf{x})$, which are themselves functions on $\Sigma $. The poisson bracket of the smeared Gauss constraint with the gauge potential thus becomes
\begin{align*}
    \{G_\lambda,A_i(\textbf{x})\}&=-\frac{\delta G_\lambda}{\delta E^i(\textbf{x})}=-\frac{\delta}{\delta E^i(\textbf{x})}\int_\Sigma  d^3y\,\lambda(\textbf{y})(\nabla\cdot\textbf{E}(\textbf{y})-\rho)=-\frac{\delta}{\delta E^i(\textbf{x})}\int_\Sigma  d^3y\,\lambda(\textbf{y})\nabla\cdot\textbf{E}(\textbf{y})
    \\
    &=-\frac{\delta}{\delta E^i(\textbf{x})}\left(\int_{\partial\Sigma}  d^2y\,\lambda(\textbf{y})\textbf{E}(\textbf{y})\cdot\textbf{n}-\int_\Sigma d^3y\,\nabla\lambda(\textbf{y})\cdot\textbf{E}(\textbf{y})\right)
    \\
    &=\partial_i\lambda(\textbf{x})-\frac{\delta}{\delta E^i(\textbf{x})}\int_{\partial\Sigma}  d^2y\,\lambda(\textbf{y})\textbf{E}(\textbf{y})\cdot\textbf{n}\,.
\end{align*}
Here, we have performed partial integration and denoted the unit normal vector pointing out of the boundary $\partial\Sigma$ by $\textbf{n}$. This boundary term should be viewed as an integral over a sphere of radius $r$, taking $r\to\infty$, since it is an asymptotic boundary. But from this derivation it becomes clear that, if the Poisson bracket $\{G_\lambda,A_i(\textbf{x})\}$ indeed is to generate gauge transformations, then the boundary term must vanish. In order to guarantee this we require that $\lambda\to 0$ asymptotically, which is how we obtain the result that redundant gauge transformations should become trivial at infinity. For a more mathematical treatment and for detailed considerations concerning the rate at which (gauge) fields and gauge parameters should vanish asymptotically, we refer to \autocite{rielloHamiltonianGaugeTheory2024,borsboom,borsboominstantaneous}.

Furthermore, it is not difficult to see that the following Poisson bracket holds between the smeared Gauss constraint and the electric field:
\begin{align*}
    \{G_\lambda,E^i(\textbf{x})\}=\frac{\delta}{\delta A_i(\textbf{x})}\int_\Sigma  d^3y\,\lambda(\textbf{y})(\nabla\cdot\textbf{E}-\rho)=0\,.
\end{align*}
Thus, the smeared Gauss constraint $G_\lambda$ does not modify the electric field: which is observable, and therefore gauge-invariant.

\subsection{Deriving Coulomb gauge}\label{coulombgauge}

The Gauss law states that the divergence of the electric field equals the distribution of charges in space. We will now discuss how the general solution of this constraint equation gives, in a natural way, what is usually called the `Coulomb gauge'. To this end, we will first solve the Gauss constraint for the electric field, $\bf E$. Since, as shown in Eq.~\eqref{pii}, the electric field is the canonical momentum associated to the gauge potential $\bf A$, the solution that we write down has to be compatible with the canonical Poisson bracket. This requirement will naturally give us the Coulomb gauge.

The most general solution \textit{in vacuo} of the Gauss constraint, which is a first-order linear differential equation, is given, in the familiar way, by a linear combination of: (i) the {\it general solution of the homogeneous equation}, i.e.~$\nabla\cdot{\bf E}=0$ (i.e.~the \textit{transverse} part of the electric field, whose divergence is zero but whose curl is non-zero), and (ii) a {\it particular solution of the inhomogeneous equation} (i.e.~the \textit{Coulombic} part of the electric field). In other words, we seek the \textit{Helmholtz decomposition} $E^i=E_L^i+E_T^i$ of the electric field into longitudinal (irrotational, i.e.~curl-free) and transverse (solenoidal, i.e.~divergence-free) components.\endnote{One could change this decomposition by adding a gradient $\nabla f$ to the longitudinal part of the electric field $\textbf{E}^L$. After all, the curl of a gradient is still zero. Thus, the Helmholtz decomposition is in this sense not unique. Adding such a gradient $\nabla f$ would change the Coulomb gauge condition as it is derived in this Section from $\nabla\cdot\textbf{A}=0$ to $\nabla\cdot\textbf{A}=\nabla f$. However, it seems most natural to simply put $\nabla\cdot\textbf{E}=\rho$, where $\rho$ is the charge density, and this gives the usual expression if $\rho=0$. We thank Ward Struyve for pointing out this non-uniqueness in the Coulomb gauge.} Since the longitudinal part $E_L^i$ is curl-free, it can be written as the gradient of a scalar function (Poincaré lemma), i.e.~$E_L^i=\partial^i \epsilon$ with $\epsilon\in C^\infty_I(\Sigma )$.\endnote{We recall that the subscript $I$ in $C^\infty_I(\Sigma)$ signifies appropriate asymptotic fall-off behaviour, a notation inherited from the group $\mathcal{G}^I$ of asymptotically constant gauge transformations. In this case we need the functions $\epsilon\in C^\infty_I(\Sigma)$ to be such that their derivatives $\partial^i\epsilon$ exhibit the same asymptotic behaviour as the longitudinal electric fields $E^i_L$.} We then generate the Coulombic electric field coordinates in phase space by the vector field
\begin{align*}
    \mathbb{E}^L=\int_\Sigma  d^3x\, E^L_i\,\frac{\delta}{\delta E_i(\textbf{x})}=\int_\Sigma  d^3x\, \partial_i\epsilon\, \frac{\delta}{\delta E_i(\textbf{x})}\,,
\end{align*}
where, following \autocite{gomesQuasilocalDegreesFreedom2021,gomesHowChooseGauge2024}, we have again used the double struck notation to stress that this vector field lives on infinite-dimensional phase space. The analogy here is with a vector $v$ in usual finite-dimensional tangent space, which can be written in a coordinate basis as $v=v^i \partial/\partial x^i$. Similarly $\delta/{\delta E_i(\textbf{x})}$ is the basis vector field in infinite-dimensional tangent space. Additionally we have to integrate over $\Sigma $, since $\delta/\delta E_i(\textbf{x})$ is a functional derivative (it derives with respect to a field which itself is a function on $\Sigma$). A vector field on the infinite-dimensional phase space should be thought of as being tangent to a curve through the space of fields $({\bf A},{\bf E})$.

The last step is to require that the longitudinal component of the canonical momentum (which is ~$\bf E$, see Eq.~\eqref{pii}) has zero Poisson bracket with any component of the gauge potential ~$\bf A$. Thus we extend the Coulombic-radiative split to the vector potential $\textbf{A}$, i.e.~we look for the component $\textbf{A}^T$ of $\textbf{A}$ that is symplectically orthogonal to $\textbf{E}^L$. To find this radiative component, we define another vector field in phase space (orthogonal to the previous one):
\begin{align*}
    \mathbb{A}^T=\int_\Sigma  d^3x~A^T_i\,\frac{\delta}{\delta A_i(\textbf{x})}\,,
\end{align*}
and we require that the two are orthogonal with respect to the symplectic structure on the phase space (see Eq.~\eqref{symplecticformmaxwell}):
\begin{align}\label{symplecticcomplementelectric}
    0&=\Omega(\mathbb{A}^T,\mathbb{E}^L)=\int_\Sigma  d^3x\, \mathbb{d}A_i\wedge\mathbb{d}E^i(\mathbb{A}^T,\mathbb{E}^L)=\int_\Sigma  d^3x \left(\mathbb{A}^T(A_i)\mathbb{E}^L(E^i)-\mathbb{E}^L(A_i)\mathbb{A}^T(E^i)\right)
    \\
    &=\int_\Sigma  d^3x\, \mathbb{A}^T(A_i)\mathbb{E}^L(E^i)=\int_\Sigma  d^3x\, A^T_i(E^L)^i=\int_\Sigma  d^3x\,A^T_i\partial^i\epsilon=-\int_\Sigma  d^3x\,\epsilon\partial^i\,A^T_i 
\end{align}
for all $\epsilon\in C^\infty_I(\Sigma )$, again using partial integration and assuming the boundary term to vanish due to sufficiently rapid asymptotic fall-off behaviour of $\epsilon$ and $A_i$. This equation must hold for any $\epsilon\in C^\infty_I(\Sigma )$, so we find $\partial^i A^T_i=0$, which, as we announced, is the Coulomb gauge condition. Thus the Coulomb gauge is singled out by the Hamiltonian formalism as the part of the gauge potential that is transverse, i.e.~orthogonal to the longitudinal part of the electric field.

We can project the gauge field $\textbf{A}$ onto the component satisfying this condition. This is called the \textit{radiative projection}, and it is given by \autocite{gomesHowChooseGauge2024}:
\begin{align}\label{radiativeprojection}
    A^T_i(\textbf{A})=A_i-\partial_i(\Delta^{-1}\partial^jA_j)\,,
\end{align}
where $\Delta^{-1}=\nabla^{-2}$ is the inverse of the Laplacian with Green's function $-\frac{1}{4\pi r}$. The radiatively projected vector potential is gauge-invariant: under a gauge transformation $A_i\to A_i+\partial_i\lambda$, we have
\begin{align*}
    A^T_i&\to A_i+\partial_i\lambda-\partial_i(\Delta^{-1}(\partial^jA_j+\partial^j\partial_j\lambda))=A_i+\partial_i\lambda-\partial_i(\Delta^{-1}\partial^jA_j)-\partial_i\Delta^{-1}\Delta\lambda
    \\
    &=A_i+\partial_i\lambda-\partial_i(\Delta^{-1}\partial^jA_j)-\partial_i\lambda=A_i-\partial_i(\Delta^{-1}\partial^jA_j)=A^T_i\,.
\end{align*}
This was certainly to be expected, since we removed precisely the part of the gauge field that is symplectically orthogonal to the part of the electric field that satisfies the constraint (the Gauss law). In other words, we removed the part of the gauge field forming a null direction of the symplectic form on the constraint surface, i.e.~the \textit{pure gauge} part in the sense of the constrained Hamiltonian formalism.

In summary, we see that the Coulomb gauge is singled out by two ingredients. First, a foliation of spacetime into spatial slices (here the $3+1$ split $\Sigma \times \mathbb{R}$) is required in order to define the 
canonical pair $(\mathbf{A}, \mathbf{E})$ on a Cauchy surface and to make sense of the Gauss law as a constraint at a fixed time. Second, given this foliation, the Coulomb gauge is the unique gauge condition that arises from decomposing the symplectic form $\Omega$ into two canonical pairs: a radiative pair $(A^T_i, E^i_T)$, whose dynamics is governed by the propagating degrees of freedom, and a Coulombic pair $(A^L_i, E^i_L)$, where the longitudinal electric field $E^i_L = \partial^i \epsilon$ is determined instantaneously by the charge distribution via $\nabla \cdot \mathbf{E} = \rho$. The condition $\partial^i A^T_i = 0$ arises from requiring these two pairs to be symplectically orthogonal with respect to $\Omega$.

However, it follows from Eq. \eqref{radiativeprojection} that the Coulomb gauge is non-local, since one needs to integrate $\textbf{A}$ together with the Green's function $-\frac{1}{4\pi r}$ over all of $\Sigma$ to obtain $A_i^T$. This non-locality might be perceived as a weakness, as compared to \textit{unitary gauge}, which is traditionally used by physicists \autocite{Weinberg:1973ew} and argued for in \autocite{wallaceGaugeInvarianceGauge2024}. However, unitary gauge can only be constructed on the assumption that the Higgs field vanishes nowhere (the same holds for the dressing field considered in Section \ref{DFM}, which is analogous to unitary gauge). Thus it can only be used to describe symmetry-broken field configurations (and even then it can only describe small fluctuations of the Higgs field), as recognized in \autocite{wallaceGaugeInvarianceGauge2024}. Coulomb gauge, on the other hand, can be used for both unbroken and broken field configurations, which makes it suitable for studying symmetry \textit{breaking}. In fact, as we will see in Section \ref{higgsqft}, Coulomb gauge is the only appropriate gauge for studying SSB in the Abelian Higgs model in QFT.

\section{Resolving the Incompatibility}\label{coulomb}

In the previous Sections, we used constrained Hamiltonian analysis to illustrate how, in electromagnetism, the Coulomb gauge allows one to remove the gauge symmetries generated by the Gauss law constraint, i.e. the redundant gauge symmetries. This has set us up to address the incompatibility, introduced in Section \ref{intro}, between the accounts of the Abelian Higgs mechanism: 

(i)~~ in terms of global gauge symmetry breaking, as in Section 7.2 of \autocite{struyveGaugeInvariantAccounts2011a}, and 

(ii)~~without symmetry breaking, as in \autocite[Section 4.2]{struyveGaugeInvariantAccounts2011a} and \autocite{berghoferGaugeSymmetriesSymmetry2023}.

The main point we endeavor to explain is that, to resolve this incompatibility, the use of the DFM to study the Higgs mechanism, i.e.~(ii), as it is done in e.g. \autocite{francoisArtificialSubstantialGauge2019,berghoferGaugeSymmetriesSymmetry2023}, is inadequate. For one cannot cover the full configuration space of classical fields by considering gauge-invariant composite objects like in these applications of the DFM. Indeed, one should not try to give a gauge-invariant account of the Higgs mechanism by eliminating the whole gauge group. Rather, one must apply the DFM at the higher level of infinite-dimensional phase space,\endnote{The details of this approach can be found in \autocite{gomesGaugingBoundaryFieldspace2019,gomesUnifiedGeometricFramework2019,gomesQuasilocalDegreesFreedom2021,rielloHamiltonianGaugeTheory2024}.} a method also known as the field-dependent DFM \autocite{Francois:2024rdm}, in order to create fields in the Coulomb gauge that are nearly gauge-invariant, i.e.~invariant under all but the global gauge transformations. This highlights how one element from Section \ref{electrogauss} is crucial for the correct application of the DFM: namely, leaving the residual global gauge symmetry in the Coulomb gauge.

The structure of this Section therefore mirrors our two points, (i) and (ii), in the Introduction: Section \ref{Struyve} first reviews the gauge-invariant account of the Higgs mechanism from \autocite[Section 7.2]{struyveGaugeInvariantAccounts2011a}. Then, Section \ref{DFM} gives the correct version of the DFM gauge-invariant account. With this correction, (i) and (ii) are rendered compatible.

\subsection{The Higgs mechanism in terms of global gauge symmetry breaking}\label{Struyve}

As we discussed in Section \ref{electrogauss}, the Coulomb gauge condition $\partial^i A_i=0$, which is enforced by the radiative projection
\begin{align}\label{radiativeproj}
    A^T_i(\textbf{A})=A_i-\partial_i(\Delta^{-1}\partial^jA_j)\,,
\end{align}
renders the gauge field invariant under the total gauge group $\mathcal{G}$, and in particular under the subgroup $\mathcal{G}^I$ of gauge transformations that are constant asymptotically, so that they preserve $\textbf{A}$'s boundary conditions. This seems to suggest that the entire discussion above was unnecessary: the Coulomb gauge simply removes the entire gauge group, as any proper gauge should. Why then the hassle of distinguishing between physical (global) and unphysical gauge transformations?

Crucially, however, we have not said anything about matter fields. In the Abelian Higgs model, we consider a complex scalar field $\varphi$, understood to be the local expression of a section of an associated vector bundle. Under a gauge transformation with parameter $\lambda$, this field transforms as:
\begin{align}\label{phit}
    \varphi\to e^{ie\lambda}\varphi\,.
\end{align}
There is something noteworthy about this transformation: the field clearly also changes under a global gauge transformation $g=e^{i\lambda_0}$ with $\lambda_0\in\R$ constant. This is not the case for the gauge field, because of the derivative in $A_i\to A_i+\partial_i\lambda$. Thus, although the global gauge group can be abstractly understood as the group of gauge transformations that have DES, its action can only be seen in the presence of matter fields. Indeed, it is the fact that the global gauge group acts trivially on gauge fields but non-trivially on matter fields that is used in the construction of Galileo's ship scenarios for $U(1)$ gauge theories in thought experiments of the `t Hooft beam splitter sort \autocite{greavesEmpiricalConsequencesSymmetries2014}. This fact is also what makes the issue of \textit{gluing} subregions in gauge theories non-trivial, and leads to Theorem \ref{gomestheorem}  \autocite{gomesQuasilocalDegreesFreedom2021,gomesHolismEmpiricalSignificance2021}.

Thus we need to examine how, in the Abelian Higgs model, the matter field can be made gauge-invariant too, in an appropriate sense. To this end, we use a redefinition similar to the radiative projection. Namely, we will dress the Higgs field with a Coulomb tail, as follows:
\begin{align}\label{phip}
    \varphi\mapsto \varphi'= e^{-ie\Delta^{-1}\partial^iA_i}\varphi\,,
\end{align}
and we will always work in terms of $\varphi'$, which as we will now argue has the right properties under gauge transformations. Indeed, the thus dressed Higgs field, Eq.~\eqref{phip}, is invariant under all gauge transformations \textit{except the global ones}, precisely because the global gauge transformations do not change $A_i$. For consider a {\it redundant}, i.e.~local but not global, gauge transformation. Then we see that the transformation of the original Higgs field $\varphi$, in Eq.~\eqref{phit}, is precisely cancelled by the transformation coming from the Coulomb dressing by the gauge field. By contrast, if the transformation is global (rigid), then the dressing does not transform, and the dressed Higgs field $\varphi'$, transforms just like $\varphi$. Thus $\varphi'$ is invariant under all gauge transformations, except the global ones.

In effect, the Coulomb dressed field $\varphi'$ in Eq.~\eqref{phip} defines equivalence classes on the space of fields: namely, the orbits of the Higgs field under {\it redundant} gauge transformations. Thus by working only with the Coulomb dressed field $\varphi'$, we in effect {\it reduce} the space of fields. Indeed, the thus reduced space of fields corresponds to the Marsden-Weinstein symplectic reduction by $\mathcal{G}^\infty_0$ \autocite{Marsden:1974dsb,Diez:2024dts}, which can be performed precisely because $\mathcal{G}^\infty_0$ acts freely on the space of fields when a Higgs field is included.

By using the radiatively projected $A^T_i$ and the redefined Higgs field $\varphi'$, Struyve is able to reformulate the Abelian Higgs mechanism gauge-invariantly. Since the only symmetry that is left when using these fields is the global gauge group, the only possible notion of SSB is global gauge symmetry breaking. Section 7.2 of \autocite{struyveGaugeInvariantAccounts2011a} then works out that all the usual consequences of the classical Abelian Higgs mechanism can indeed be derived from {\it global gauge symmetry breaking}.\\
\\
{\it Interpretation of SSB in the gauge-invariant Higgs mechanism.} In view of our discussion on global gauge symmetries as the physical asymptotic gauge group, we believe that the above is the correct way to understand the Higgs mechanism. For it is a \textit{gauge-invariant account} of the Higgs mechanism, in the correct sense: namely, it is invariant under {\it redundant} gauge symmetries, viz.~the ones that are local but not global. And it is not, and it should not, be invariant under those symmetries that have DES. For, as we have discussed, those symmetries are relationally physical in the sense that they correspond to different configurations of the physical subsystem (here, the Higgs field) when viewed from a(n) (asymptotic) boundary. Thus spontaneously breaking the global symmetry means choosing a specific state of the Higgs field, out of a number of possible states. Namely, {\it the symmetry breaking in the Higgs mechanism is the specification of the physical state of the Higgs field}, and this state is invariant under the redundant gauge symmetries. On Minkowski spacetime one can think of this state as a specification of an asymptotic condition for the field, which determines the global phase value that the Higgs field must approach asymptotically (Of course, we have not yet said anything about the other aspect of the Higgs mechanism, beyond symmetry breaking: namely, the generation of mass. We will return to this in Section \ref{higgsqft}).

\subsection{The correct gauge-invariant DFM}\label{DFM}

In this light, we can finally resolve the tension between Struyve's account and the dressing field method as presented in \autocite{massonRemarkSpontaneousSymmetry2010,francoisArtificialSubstantialGauge2019,berghoferGaugeSymmetriesSymmetry2023}. Let us briefly review the main idea of the DFM as presented there: considering a principal $G$-bundle $\pi\colon P\to \Sigma$, and letting $H\subset G$ denote a closed subgroup, the fundamental object of the DFM is:
\begin{definition}
    A smooth map $u\colon P\to H$ satisfying $u(ph)=h^{-1}u(p)$ for all $h\in H$ is called an $H$-\textit{dressing field}.
\end{definition}
We can think of dressing fields as trivializations in the direction of the subgroup $H$, as formalized in the following result \autocite{fournelGaugeInvariantComposite2014}.
\begin{proposition}\label{dressingfieldprop}
    A dressing field $u\colon P\to H$ exists if and only if there is an isomorphism of $H$-spaces $P\cong P/H\times H$, where the action of $H$ on $P/H$ is trivial.
\end{proposition}
Thus, a dressing field partially trivalizes the principal bundle. It can then be used to dress gauge and matter fields in order to make them invariant under $H$-valued gauge transformations.
\begin{definition}
    Let $u$ be an $H$-dressing field and $A\in\Omega^1(P,\mathfrak{g})$ a connection 1-form with curvature $F$. Let $\rho\colon G\to V$ be a representation giving an associated bundle $E=P\times_\rho V$ and $\varphi\colon P\to V$ a $G$-equivariant map (equivalently a section of $E$). Then we define the \textit{dressed fields}
    \begin{align*}
        A^u&=u^{-1}Au+u^{-1}du,
        \\
        \varphi^u&=\rho(u^{-1})\varphi,
        \\
        F^u&=u^{-1}Fu=dA^u+\frac{1}{2}[A^u,A^u],
        \\
        D^u\varphi^u&=\rho(u^{-1})D\varphi=d\varphi^u+\rho_*(A^u)\varphi^u.
    \end{align*}
    We note that $A^u$ is not itself a connection 1-form, which should not be expected anyway since $u$ is not a gauge transformation.
\end{definition}
\noindent{\it Critique of the standard use of DFM.} In \autocite{francoisArtificialSubstantialGauge2019,berghoferGaugeSymmetriesSymmetry2023} the DFM is used to distinguish between \textit{artificial} and \textit{substantial} gauge symmetries, i.e.~those for which a local dressing field can be found and those for which only a non-local dressing field can be obtained. This way, the well-known trade-off between locality and gauge-invariance, which is famously showcased by the Aharonov-Bohm effect \autocite{healeyGaugingWhatReal2007}, is formalized. When applied to the Abelian Higgs model, the DFM then supposedly shows that the $U(1)$ gauge symmetry is artifical, since one can define a dressing field $u$ through the polar decomposition $\varphi=u\sqrt{\varphi^*\varphi}$ of the Higgs field \autocite{berghoferGaugeSymmetriesSymmetry2023}, in analogy to unitary gauge. Something similar is done to remove the $SU(2)$ symmetry of the electroweak $U(1)\times SU(2)$ theory. It is then concluded that, since there is no gauge symmetry left in the theory, there can be no SSB in the Higgs mechanism. This leads to the idea that the DFM provides ``an alternative interpretation of the BEHGHK mechanism that
is more in line with the conclusions of the community of philosophers of physics'' \autocite[p. 4]{attardDressingFieldMethod2018}. In particular ``the DFM approach to the electroweak
model is consistent with Elitzur’s theorem stating that in lattice
gauge theory a gauge symmetry cannot be spontaneously broken'' \autocite[p. 66]{berghoferGaugeSymmetriesSymmetry2023}. François' article even contains a section titled ``there is no SSB in the electroweak model and we long suspected it'' \autocite{francoisArtificialSubstantialGauge2019}.

However, as noted also in \autocite{berghoferGaugeSymmetriesSymmetry2023}, there is something awkward about the application of the DFM to the Abelian Higgs mechanism or electroweak theory: the polar decomposition of the Higgs field is not defined when the Higgs field vanishes. Berghofer et al. suggest simply ignoring the vanishing field configurations, but we are now in a position to understand its deeper sense. The gauge group $\mathcal{G}^I$ does not act freely on the space of $U(1)$ gauge fields, because of the global gauge transformations. This means that one cannot take a smooth symplectic quotient\endnote{Instead one obtains a stratified space with strata labelled by the electric flux \autocite{rielloHamiltonianGaugeTheory2024}.} - one cannot properly parametrize the entire field space in terms of gauge-invariant variables, as is attempted in \autocite{berghoferGaugeSymmetriesSymmetry2023}. Yet, when the Higgs field is added, the issue is apparently resolved, since the gauge group does act freely on the space of Higgs fields. After all, global gauge transformations do change the Higgs field by a phase factor. Naively, then, we can use the phase of the Higgs field as a reference frame. But unfortunately, the problem remains, since the configurations $(\varphi=0,A_i)$ of vanishing Higgs field and arbitrary gauge field are still invariant under the action of the global gauge group. Thus, even on the enlarged phase space of gauge and Higgs fields one cannot perform smooth symplectic reduction. It should not come as a surprise that, if one ignores vanishing Higgs field configurations, such a symplectic quotient seems possible. After all, it is precisely these configurations that are problematic. But there is no legitimate reason to remove vanishing Higgs field configurations. Indeed, in doing so one removes the ``original'' symmetric state of the Higgs field ``before'' symmetry breaking, so it is not surprising that one does not find any SSB in this approach. Put differently, if one wants to model the unbroken phase of the Higgs model, then all Higgs fields are required to approach zero asymptotically. So in that case the set of Higgs fields that vanish somewhere actually has measure 1. Neglecting the somewhere-vanishing Higgs configurations therefore amounts to an \textit{a priori} exclusion of the possibility of modeling the unbroken phase of the model.

What, then, is left of the DFM? Should we dismiss it entirely? Not at all, but we must apply it at a different level, in the spirit of \autocite{gomesUnifiedGeometricFramework2018,gomesUnifiedGeometricFramework2019} and of the field-dependent DFM \autocite{Francois:2020tom,Francois:2021jrk,Francois:2024rdm}. We should not attempt to use it to remove, for instance, the $SU(2)$ part of the structure group of the electroweak theory. Instead, we should implement it to remove the entire \textit{unphysical} subgroup $\mathcal{G}^\infty_0$ (which equals $\mathcal{G}^\infty$ for $G=U(1)$). In other words, we should seek a dressing field \textit{at the level of infinite-dimensional field space}, i.e.~an appropriately transforming $\mathcal{G}^\infty_0$-valued map on field space. But, for $G=U(1)$, this is precisely provided by the object $\text{exp}(-i\Delta^{-1}\partial^i A_i)$, viewed as a map on field space! Indeed, it is not hard to see that for any gauge transformation $e^{i\lambda}\in\mathcal{G}^\infty$ we have
\begin{align*}
    e^{-i\Delta^{-1}\partial^i A_i}\to e^{-i\Delta^{-1}\partial^i(A_i+\partial_i\lambda)}= e^{-i\lambda}e^{-ie\Delta^{-1}\partial^i A_i}\,,
\end{align*}
as required in the definition of a dressing field. In addition, if we indeed take the field-dependent dressing field $u=\text{exp}(-i\Delta^{-1}\partial^i A_i)$ on field phase space, the ``pure gauge term'' $u^{-1}\mathbb{d}u$ reproduces the radiative projection. Thus, we see that at the higer level of infinite-dimensional field space, the DFM and Struyve's account actually harmonize perfectly. The Coulomb gauge is reproduced by the dressing field $\text{exp}(-i\Delta^{-1}\partial^i A_i)$, and the two approaches conspire to reveal the true sense of a ``gauge-invariant account'' of the Higgs mechanism as one in which the unphysical gauge group $\mathcal{G}^\infty$ is removed, so that the focus lies entirely on the breaking of the physical global gauge group $\mathcal{G}^I/\mathcal{G}^\infty\cong U(1)$.

\section{Abelian Higgs Mechanism in QED}\label{higgsqft}

Pleasing as the above solution of our first research question may be, the relation to quantum theory remains entirely unilluminated. The goal of this Section is to show that the classical ideas presented so far extend to QFT. Indeed, Morchio and Strocchi\endnote{These are also the M and S in the FMS approach \autocite{frohlichHiggsPhenomenonSymmetry1980,frohlichHiggsPhenomenonSymmetry1981}, which is referred to in \autocite{berghoferGaugeSymmetriesSymmetry2023} as a quantum-theoretical extension of the DFM. But the FMS approach is perturbative and only works on the assumption that the Higgs field is nowhere vanishing. The results of Morchio and Strocchi showcased in this Section are non-perturbative.} \autocite{morchioLocalizationSymmetries2007,strocchiIntroductionNonPerturbativeFoundations2016,strocchiSymmetryBreakingStandard2019} have shown that mass generation in the Abelian Higgs mechanism can be understood through SSB of global $U(1)$ gauge symmetry \textit{in the algebraic sense}, answering (though probably unknowingly) the original call to do so in \autocite{earmanRoughGuideSpontaneous2003} (which was repeated and refined in \autocite{stoltznerConstrainingHiggsMechanism2012}). 

We begin by recalling the definition of SSB in algebraic quantum theory. We then present the Morchio and Strocchi's main theorem, which uses this algebraic language to show that there can only be massless photons in scalar QED if the global $U(1)$ gauge symmetry is \textit{unbroken}, in which case the electric charge is \textit{superselected}. If the $U(1)$ symmetry is \textit{broken}, we do not have superselection but instead we have \textit{charge screening}, and the photons must be massive. The core result (Theorem \ref{strocchiabelianhiggscoulomb}) is fairly complicated and technical, so we spend some time preparing for its exposition and explaining its various parts. We note that all relevant proofs are neatly collected in Chapters 5 and 6 of \autocite{borsboom2024spontaneous}.

\subsection{SSB in algebraic quantum theory}\label{quantumssb}

Let us recall one of the two equivalent definitions of SSB in the algebraic language of $C^*$-algebras \autocite[p. 345]{landsmanFoundationsQuantumTheory2017} (we refer to \autocite{borsboomdupont} for a conceptual study of various proposed definitions of SSB).
\begin{definition}\label{ssbdefn}
    Let $\mathcal{A}$ be a $C^*$-algebra and $\beta\colon \mathcal{A}\to\mathcal{A}$ a $*$-automorphism. Then a state $\omega\colon \mathcal{A}\to\C$ is said to spontaneously break the symmetry $\beta$ if the GNS presentations $\pi_\omega\colon\mathcal{A}\to B(H_\omega)$ and $\pi_{\beta^*\omega}\colon\mathcal{A}\to B(H_{\beta^*\omega})$ are not unitarily equivalent, i.e.~if there is no unitary operator $U\colon H_\omega\to H_{\beta^*\omega}$ such that
    \begin{align*}
        \pi_{\beta^*\omega}(a)=U\pi_\omega(a)U^*,\;\;\;\;\; a\in\mathcal{A}\,.
    \end{align*}
    Here the state $\beta^*\omega$ is defined by $\beta^*\omega(a)=\omega(\beta^{-1}(a))$.
\end{definition}
Now, this definition of SSB is simple and general, but not so easy to check \autocite[p. 120]{strocchiSymmetryBreaking2008}. We should like to make use of an \textit{order parameter}, i.e.~an element of the algebra whose ground state expectation value is not invariant under the symmetry in question precisely when that symmetry is broken. This is how we will study the Higgs mechanism in Section \ref{higgsmehcanism}, and to that end we have the following result (cf. Proposition II.8.2 in \autocite{strocchiSymmetryBreaking2008}). For the proof the reader can use Strocchi's original work, but a more pedagogical treatment is provided under Proposition 5.26 in \autocite{borsboom2024spontaneous}.

\begin{proposition}{\bf (Symmetry breaking).}\label{orderparameterprop}
    Let $\mathcal{A}$ be a $C^*$-algebra with vacuum state $\omega_0\in S(\mathcal{A})$ in whose GNS representation $(\pi_{\omega_0},H_{\omega_0},\Omega_{\omega_0})$ spacetime translations $\beta_x$ are implementable through a strongly continuous family of unitary operators $U(x)$, such that $\Omega_0$ is the unique translationally invariant state in $H_{\omega_0}$. Then an \textit{internal} symmetry (one that commutes with spacetime translations) $\beta\in\text{Aut}(\mathcal{A})$ is unbroken in $\omega_0$, in the sense of Definition \ref{ssbdefn}, if and only if all ground state correlation functions are invariant under $\beta$, i.e.~for all $a\in\mathcal{A}$,
    \begin{align}\label{symmimplcond}
        \omega_0(\beta(a))=\omega_0(a)\,.
    \end{align}
\end{proposition}

In other words: SSB of an internal symmetry by the ground state can be detected by means of an order parameter, i.e.~an element $a\in\mathcal{A}$ whose ground state expectation value is not invariant under the symmetry transformation $\beta$. The condition in Eq.~\eqref{symmimplcond} can also be made infinitesimal:
\begin{align}\label{orderparaminfinitesimal}
    \omega_0(\delta a)=0 \;\;\;\;\;\text{  for all  } a\in\text{Dom}(\delta)\,.
\end{align}
Here $\delta$ is the derivation on $\mathcal{A}$ induced by the 1-parameter group of automorphisms $\beta_\lambda$, i.e.
\begin{align*}
    \delta(a)=\frac{d \beta_\lambda(a)}{d\lambda}\bigg|_{\lambda=0}\,,
\end{align*}
where $\beta_\lambda=\beta(\lambda)$ and $\text{Dom}(\delta)$ consists of all elements of $\mathcal{A}$ for which this limit exists (cf. Proposition 9.19 in \autocite{landsmanFoundationsQuantumTheory2017}). We note that the infinitesimal version of the symmetry implementability condition, i.e. Eq.~\eqref{orderparaminfinitesimal}, is also used in the algebraic version of Goldstone's theorem \autocite{kastlerConservedCurrentsAssociated1966,buchholzNewLookGoldstone1992,stoltznerConstrainingHiggsMechanism2012} (cf. Theorem III.3.27 in \autocite{haagLocalQuantumPhysics1996}).

Now that we have a concrete test of SSB of an internal symmetry by the ground state, we can apply this to the Abelian Higgs mechanism. Indeed, gauge symmetries are by definition internal, so we can detect global $U(1)$ gauge symmetry breaking in the vacuum state by means of an order parameter. When doing so, we do not have to worry about Elitzur's theorem, since that theorem applies only to \textit{local} gauge symmetries and not to global ones \autocite{strocchiIntroductionNonPerturbativeFoundations2016}.

\subsection{The Higgs mechanism as quantum global gauge symmetry breaking}\label{higgsmehcanism}

In this Section we discuss non-perturbative results based mostly on \autocite{morchioLocalizationSymmetries2007,depalmaNonperturbativeArgumentNonabelian2013,strocchiIntroductionNonPerturbativeFoundations2016,strocchiSymmetryBreakingStandard2019}, that are formulated in the language of the Wightman axioms of QFT, in which fields $\varphi_i(x)$ are modeled as operator-valued tempered distributions.\endnote{For details on these axioms, we refer the reader to \autocite{wightmanFIELDSOPERATORVALUEDDISTRIBUTIONS1965,streaterPCTSpinStatistics2016}.} Such tempered distributions ``eat'' test functions $f\in\mathcal{S}(\R^4)$ in Schwarz space to produce operators on a common dense domain $D\subset H$ of a Hilbert space $H$, containing the vacuum state $\Psi_0$. To a region $\mathcal{O}\subset M$ of Minkowski spacetime, we then associate the algebra generated by all polynomials of smeared fields $\varphi_i(f)$ with $\text{supp}(f)\subset\mathcal{O}$. Such smeared fields behave almost like operator-valued functions,\endnote{Technically, however, the fields are not quite well-defined \textit{at a point}.} which is why fields are written $\varphi_i(x)$ and the machinery of classical gauge fields (like Noether's theorems) can straightforwardly be imported to QFT \autocite{streaterPCTSpinStatistics2016}. We note that the Wightman recipe of producing local algebras by smearing with locally supported test functions does not necessarily yield $C^*$-algebras, and indeed the relation between the Haag-Kastler and Wightman axioms is subtle \autocite{haagLocalQuantumPhysics1996}. We assume, however, that the detection of SSB via an order parameter as in Proposition \ref{orderparameterprop} is still possible for the Coulomb field algebra of QED considered in this Section.

When quantizing a gauge field theory in terms of the Wightman axioms, various approaches exist. A symmetry transformation is an automorphism of the total algebra of fields operators, but it can also be viewed as a transformation of the states. The main dichotomy in gauge quantization is about the question of what to do with the Gauss law constraint. One could either impose it on the smeared operator-valued fields, i.e. as an ``operator equation'', or one could decide to impose it only later on the states in the Hilbert space built from the action of the fields on the vacuum state \autocite{strocchiIntroductionNonPerturbativeFoundations2016,landsmanFoundationsQuantumTheory2017}. The latter approach is known in QED as Gupta-Bleuler quantization, which we pursue in Section \ref{noetherlocality}, pointing out its problems. To remedy these problems we will move from Gupta-Bleuler quantization to non-local Coulomb gauge quantization, in which the Gauss constraint is imposed on the fields rather than the states, leaving only a global $U(1)$ gauge symmetry. For more details on gauge symmetries in algebraic QFT and the relation to the Wightman approach pursued here, we refer to Chapters 5 and 6 of \autocite{borsboom2024spontaneous}.

The main result of this Section, Theorem \ref{strocchiabelianhiggscoulomb}, is centered around the Gauss law $\rho=j^0=\nabla\cdot\textbf{E}=\partial_ iF^{i0}$ in Coulomb gauge quanization, viewed as a consequence of Noether's second theorem. We will therefore also refer to it as the \textit{Noether relation}, following Strocchi's terminology. Through this Noether relation the electric charge is expressed as a \textit{Gauss charge}, i.e.~in terms of the antisymmetric field strength tensor $F_{\mu\nu}$. In Theorem \ref{strocchiabelianhiggscoulomb}, it will be this \textit{very relation which is at stake}, in the sense that it holds in QED only if the global $U(1)$ symmetry is unbroken. Its failure to hold when the $U(1)$ symmetry is broken by the vacuum state can be understood to be a consequence of current charge screening: if the electromagnetic force is carried by massive photons, then it is short-ranged, so electric charges cannot be detected infinitely far away by means of Gauss's theorem.

Thus, we first develop the quantum-theoretical version of the Noether relation between the electric charge and the electromagnetic field strength, which is in fact not so straightforward to carry over from classical field theory to QFT. In particular, we explain how there is a conflict between the Gauss law and locality, which forces us to allow for non-local fields in the Coulomb gauge, thereby enabling us to recover the Noether relation. Continuing in the Coulomb gauge, we then present the main theorem and explain the meaning and relevance of superselection and current charge screening.

\subsubsection{The Noether relation and locality}\label{noetherlocality}

We recall that the Gauss law arises through Noether's second theorem as the constraint for Yang-Mills theory, such that the Gauss charge, i.e.~the electric charge $Q^e=\int d^3x j_0(\textbf{x},0)$ for electromagnetism (here $j_0(\textbf{x},0)$ denotes the time-component at time $t=0$ of the electric current $j_\mu=\partial^\nu F_{\nu\mu}$), can be expressed in terms of the antisymmetric field strength tensor $F_{\mu\nu}$. By Noether's first theorem this electric charge generates (by taking Poisson brackets) global $U(1)$ gauge transformations of an electromagnetically charged scalar field $\varphi$ with current $j_\mu$:
\begin{align*}
\{Q^e,\varphi\}=\delta^e\varphi=ie\varphi\,,\;\;\;\;\; \{Q^e,A_\mu\}=\delta^e A_\mu=0\,.
\end{align*}
In other words: by Noether's theorems, we know that the integral $\int d^3x\, j_0(\textbf{x},0)=\int d^3x\, \partial^i F_{i0}(\textbf{x},0)$ generates global $U(1)$ transformations. This fact should be extended to QFT by replacing Poisson brackets with commutators in order to obtain a well-defined notion of a quantum scalar field charged under the electromagnetic interaction, with a global $U(1)$ Noether symmetry corresponding to this charge. However, one runs into problems when quantizing while imposing locality, as we will now see.

Let us attempt to extend the generation of global gauge  transformations by the electric charge to quantum fields $\varphi(x),A_\mu(x)$ assumed to satisfy the Wightman axioms in the local Gupta-Bleuler quantization \autocite{strocchiIntroductionNonPerturbativeFoundations2016}. We first need to regularize the charge by smearing it out in space and time:
\begin{align}\label{gausscharge}
    Q_R=\int d^3 x d t f_R(\mathbf{x}) \alpha(t) j_0(\mathbf{x}, t) \equiv j_0\left(f_R \alpha\right),
\end{align}
where $f_R(\textbf{x})=f(\textbf{x}/R)\in\mathcal{D}(\R^3)$ with $f(x)=1$ for $|\textbf{x}|<1$ and $f(\textbf{x})=0$ for $|\textbf{
x}|<1+\delta$ with $\delta>0$, and where $\text{supp}(\alpha)\subset[-\gamma,\gamma]$ with $\int\alpha(t)dt=1$ \autocite[p. 147]{strocchiIntroductionNonPerturbativeFoundations2016}. We have to perform this regularization because the limit $Q_R$ for $R\to\infty$ does not actually exist. By regularising we can avoid this problem, since we only need the limits of the commutator of $Q_R$ with the matter and gauge fields. These limits do exist and are actually independent of the choice of smearing function $\alpha$ \autocite{strocchiIntroductionNonPerturbativeFoundations2016}. It it then not too hard to show that, for a theory with a scalar field $\varphi(\textbf{x},t)$ and current $j_\mu(\textbf{x},t)$ that are local with respect to each other (i.e.~that commute at spacelike separated points), we have \autocite{strocchiIntroductionNonPerturbativeFoundations2016}:
\begin{align*}
    i \lim _{R \rightarrow \infty}\left[Q_R, \varphi(g, t)\right]=i \lim _{R \rightarrow \infty}\left[j_0\left(f_R, t\right), \varphi(g, t)\right]=\delta^e \varphi(g, t),\;\;\;\;\;g\in\mathcal{D}(\R^3)\,,
\end{align*}
where $\delta^e$ denotes the derivation corresponding to the global $U(1)$ gauge symmetry. This is indeed the desired quantum version of the classical Noether relation between the charge (the integral over the current density) and the global gauge symmetry transformation. This result, however, is spoiled in the presence of the Gauss law (Proposition 2.1 in \autocite{strocchiIntroductionNonPerturbativeFoundations2016}).
\begin{proposition}\label{noetherfail}
    If a scalar field configuration $\varphi$ is local with respect to the electric field $\textbf{E}$ (i.e.~if these fields commute at spacelike separated points), then the Gauss law gives $\lim _{R \rightarrow \infty}[j_0(f_R\alpha),\varphi]=0$, implying that $Q_R$ cannot generate the global $U(1)$ transformations.
\end{proposition}

Conversely, this proposition tells us that if $\varphi$ is charged in the sense that $\delta^e\varphi\neq 0$, then it cannot be local with respect to the electric field. Thus, if we denote by $\mathcal{F}$ the local field algebra (generated by polynomials of the fields), then charged states cannot be in the closure of $\mathcal{F}\Psi_0$ (where $\Psi_0$ is the vacuum state), but must instead be generated from the vacuum by non-local charged fields \autocite{strocchiIntroductionNonPerturbativeFoundations2016}. In algebraic language, if $\mathcal{O}$ is a bounded region of spacetime with causal complement $\mathcal{O}'$, then for sufficiently large $R$ we have $Q_R\in \mathcal{A}(\mathcal{O}')$. For any algebraic state $\omega$ which is localized in $\mathcal{O}$ we then get $\omega_0(Q_R)=\langle\Psi_0,Q_R\Psi_0\rangle=0$, showing that $\omega$ is uncharged. In other words: localized states must be uncharged because of the Gauss law. 

This conflict between locality and the Gauss law\endnote{This conflict can be seen as yet another incarnation of the familiar trade-off between locality and gauge-invariance.} forces us to accept that the field algebra becomes non-local, if we want to maintain the Gauss law as an operator equation \autocite{strocchiIntroductionNonPerturbativeFoundations2016}. This is what is done in the Coulomb gauge, which can be obtained from the local field algebra by the same field transformation as in Section \ref{coulombgauge}. That is: if $A_\mu$ and $\varphi$ are the local but gauge-dependent fields in the Gupta-Bleuler quantization of QED, then the transformation to the non-local Coulomb gauge corresponds to imposing the Gauss law constraint as an operator equation on the fields by means of the radiative projection introduced in Eq.~\eqref{radiativeprojection} and Eq.~\eqref{phip}:
\begin{align*}
    \varphi_C&=e^{-ie\Delta^{-1}\partial^i A_i}\varphi,
    \\
    A^C_\mu&=A_\mu-\partial_\mu \Delta^{-1}\partial^i A_i\,.
\end{align*}
We write $A^C_\mu$ instead of $A^T_\mu$ as before in order to stress that we are now considering a \textit{quantum} field that generates the Coulomb field algebra. The field $\varphi_C$ is still charged since it transforms under the global charge group $U(1)$, but it is gauge-invariant in the redundant sense. The cost is that the fields are non-local because of the non-local nature of the operator $\Delta^{-1}\partial^i$, which involves an integral over all of space. Of course the fields are still operator-valued tempered distributions, which can be thought of as operators localized to small regions of spacetime, but space-like separated fields now no longer commute because of the Gauss law (one can measure charges by performing a surface integral arbitrarily far away).

\subsubsection{Global gauge symmetry breaking in Coulomb gauge}

We will now explain in what sense the Noether relation between the electric charge and the electromagnetic current density, which, as we saw in Proposition \ref{noetherfail}, is spoiled for local gauge quantization, can be restored in the non-local Coulomb gauge whenever the global $U(1)$ gauge symmetry is \textit{unbroken}.\endnote{As pointed out by an anonymous referee, this raises the question of whether different gauge quantization yield different quantum theories. This is a difficult philosophical question which, to our knowledge, has not been addressed. However, we expect that it will be difficult to obtain a satisfactory answer until non-perturbative quantum gauge field theories are properly understood mathematically. For now we will take a more pragmatic stance, in which we simply view the Coulomb gauge as the appropriate gauge to study the particular phenomenon of gauge symmetry breaking in the Abelian Higgs mechanism. Other gauges might be more appropriate for other phenomena.} We then present and subsequently dissect the main result of \autocite{morchioLocalizationSymmetries2007}.

Let $\mathcal{F}_C$ denote the field algebra generated by the Coulomb fields $\varphi_C$ and $A^C_i$. We assume its vacuum correlations to be well-defined (see footnote 49 in chapter 7 of \autocite{strocchiIntroductionNonPerturbativeFoundations2016}). We call the generator of the global $U(1)$ symmetry the electric charge $Q^e$, defined by
\begin{align*}
    [Q^e,\varphi_C]=e\varphi_C,\;\;\;\; Q^e\Psi_0=0\,.
\end{align*}
The question now is whether this electric charge can indeed be written like $Q_R$ in Eq.~\eqref{gausscharge}, i.e. as an integral over the current density $j_0$. The following result (Proposition 4.1 in \autocite{morchioLocalizationSymmetries2007}, Proposition 5.3 in \autocite{strocchiIntroductionNonPerturbativeFoundations2016}) shows that the correspondence between the generator of the global $U(1)$ symmetry and the current density still fails.
\begin{proposition}\label{coulombtimedependent}
    In the Coulomb gauge of QED, with field algebra $\mathcal{F}_C$ and vacuum vector $\Psi_0$, we have that for all states $\Psi,\Phi\in\mathcal{F}_C\Psi_0$ the limit
    \begin{align*}
        \lim _{R \rightarrow \infty}(\Psi,[j_0^C(f_R\alpha),\varphi_C]\Phi)
    \end{align*}
    exists. However, the limit is generally dependent on $\alpha$, meaning that the time-independent global $U(1)$ symmetry cannot be generated by such integrals of the charge density.
\end{proposition}

This failure of the integral of the charge density to generate the time-independent global $U(1)$ transformations can be remedied by time-averaging the integral of $j_0$ with an improved smearing: we define $Q_{\delta R}=j_0(f_R\alpha_{\delta R})$, where $\alpha_{\delta R}=\alpha(x_0/\delta R)/\delta R$, with $\alpha\in\mathcal{D}(\R), 0<\delta <1$ and $\text{supp}(\alpha)\subset[-\epsilon,\epsilon]$ with $\epsilon\ll 1$ \autocite{morchioLocalizationSymmetries2007}. However, this only works in the case of \textit{unbroken} $U(1)$ symmetry, and this insight is at the heart of Theorem \ref{strocchiabelianhiggscoulomb} (which is Theorem 7.6.2 in \autocite{strocchiIntroductionNonPerturbativeFoundations2016}, Theorem 2.8.3 in \autocite{strocchiSymmetryBreakingStandard2019}).

Theorem \ref{strocchiabelianhiggscoulomb} distinguishes between two situations: situation \textbf{(A)}, where there are \textit{massless photons}, and the global $U(1)$ gauge symmetry is unbroken; and situation \textbf{(B)}, where the $U(1)$ symmetry is broken by the presence of an order parameter, and all bosons must be massive. As follows:

\begin{theorem}\label{strocchiabelianhiggscoulomb}
Let $\mathcal{F}_C$ denote the Coulomb field algebra with vacuum state $\Psi_0\in H$, generated by the gauge field $A^C_i$ and the complex scalar Higgs field $\varphi_C$. Let $\beta_\lambda$ denote the 1-parameter family of $*$-automorphisms of $\mathcal{F}_C$ corresponding to the continuous global $U(1)$ symmetry with generator $Q^e$ (the electric charge), and let $j_0$ be the associated conserved Noether charge current. Then the following results hold:
\\
\\
\noindent\textbf{(A)}~~If the spectral measure $d\rho(m^2)$ of the electromagnetic field $F_{\mu\nu}$ contains a $\delta(m^2)$ contribution, i.e.~if there are massless vector bosons, then:
\begin{quote}
(i) the global $U(1)$ gauge transformations are generated by the improved smeared current density, i.e.~for any $F\in\mathcal{F}_C$
\begin{align}\label{noetherrelationrestored}
    [Q^e,F]=\delta^e F=\frac{d\beta_\lambda(F)}{d\lambda}\bigg|_{\lambda=0}=i\lim_{\delta\to 0}\lim_{R\to\infty}[j_0(f_R\alpha_{\delta R}),F]\,;
\end{align}
(ii) we have $\lim_{R\to\infty}j_0(f_R\alpha_{\delta R})\Psi_0=0$, so that $\langle\delta^e F\rangle_0=0$ for any $F\in\mathcal{F}_C$, meaning that the global $U(1)$ symmetry is unbroken (by Proposition \ref{orderparameterprop});

(iii) the electric charge $Q^e$ can be expressed in terms of $j_0$ not only in the commutators with fields $F\in\mathcal{F}_C$, but also in the matrix elements of the Coulomb charged states $\Phi,\Psi\in\mathcal{F}_C\Psi_0$:
\begin{align*}
    \langle \Phi,Q^e\Psi\rangle =\lim_{\delta\to 0}\lim_{R\to\infty}\langle \Phi,j_0(f_R\alpha_{\delta R})\Psi\rangle\,,
\end{align*}
implying that $Q^e$ is a superselected charge (see below for an explanation).
\end{quote}
\noindent\textbf{(B)}~~If the global $U(1)$ symmetry is broken by some $F_\text{SSB}\in\mathcal{F}_C$ such that $\langle\delta^e F_\text{SSB}\rangle_0\neq 0$, then
\begin{quote}
    (i) the spectral measure $d\rho(m^2)$ of $F_{\mu\nu}$ cannot contain a $\delta(m^2)$ contribution, i.e.~there are no massless vector bosons;

    (ii) the global $U(1)$ gauge transformations are not generated by the current density, in fact
    \begin{align}\label{b21}
        \lim_{\delta\to 0}\lim_{R\to\infty}[j_0(f_R\alpha_{\delta R}),F]=0
    \end{align}
    for any $F\in\mathcal{F}_C$, and $j_0(f_R\alpha_{\delta R})$ annihilates the vacuum, so that for any $\Psi\in\mathcal{F}_C\Psi_0:$
    \begin{align}\label{b22}
        \lim_{\delta\to 0}\lim_{R\to\infty}j_0(f_R\alpha_{\delta R})\Psi=0\,,
    \end{align}
    i.e.~we have \textit{current charge screening};

    (iii) the two-point function $\langle j_\mu(x)F_\text{SSB}\rangle_0$ does not vanish, and its Fourier spectrum coincides with the energy-momentum spectrum of $F_{\mu\nu}$, so that the absence of massless vector bosons coincides with the absence of massless Goldstone modes.
\end{quote}
\end{theorem}

Let us unpack this result. As we mentioned above, it distinguishes between two situations: \textbf{(A)}, in which there are \textit{massless photons}, in the sense of a zero-mass contribution $\delta\rho(m^2)$ to the Källén-Lehmann spectral representation of the electromagnetic field $F_{\mu\nu}$, and in which the global $U(1)$ gauge symmetry is unbroken; and situation \textbf{(B)}, in which the $U(1)$ symmetry is broken in the sense of Proposition \ref{orderparameterprop} by some element $F_\text{SSB}\in\mathcal{F}_C$. In the latter situation it is derived that the spectral measure of the electromagnetic field \textit{cannot} have a zero-mass contribution, i.e.~all gauge bosons must be massive. In fact, there are not even Goldstone bosons, since their spectrum coincides with that of $F_{\mu\nu}$. This latter fact corresponds to what is sometimes called the ``eating of the Goldstone modes''.

In situation \textbf{(A)}, the Noether relation between the electric charge $Q^e$ and the improved smeared current density holds in commutators. That is: the action of the derivation $\delta^e$, corresponding to the infinitesimal global $U(1)$ gauge transformation generated by $Q^e$, on an arbitrary field $F\in\mathcal{F}_C$, can indeed be written as an appropriate limit of a commutator $[j_0(f_R\alpha_{\delta R}),F]$. Furthermore, in this situation the vacuum is annihilated by the smeared current density:
\begin{align*}
    \lim_{R\to\infty}j_0(f_R\alpha_{\delta R})\Psi_0=0\,,
\end{align*}
from which it follows by the Noether relation (Eq.~\eqref{noetherrelationrestored}) that the vacuum expectation values $\langle\delta^e F\rangle_0$ vanish\endnote{After all, in the vacuum expectation values the object $\delta^e F$ is sandwiched between two copies of the vacuum state $\Psi_0$. Since $[\delta^e F]$ can be written as a limit of the commutator $[j_0(f_R\alpha_{\delta R}),F]$, the object $\lim_{R\to\infty}j_0(f_R\alpha_{\delta R})$ will always hit $\Psi_0$ on the left or right, giving zero.} for any $F\in\mathcal{F}_C$, which is equivalent to the global $U(1)$ gauge symmetry being unbroken, by Proposition \ref{orderparameterprop}. Thirdly, the Noether relation between the electric charge and the improved smeared current holds not only in commutators but also in matrix elements between states generated from the vacuum $\Psi_0$ by the action of the field algebra $\mathcal{F}_C$. From this, it can be deduced that for any observable $A\in\mathcal{F}_C$, which must be gauge-invariant, the commutator $[Q^e,A]$ vanishes on the states. This phenomenon is known as superselection. It provides a \textit{superselection rule} for states carrying different charges: these cannot form a coherent superposition.\endnote{For a philosophical introduction to this topic, we refer the reader to \autocite{earmanSuperselectionRulesPhilosophers2008}.} There is, in fact, a much more general and profound relation between global gauge symmetries and superselection, known as \textit{DHR superselection theory} \autocite{doplicherWhyThereField1990,haagLocalQuantumPhysics1996}. Indeed, representations of field algebras with global gauge symmetry form different \textit{superselection sectors} labeled by the gauge charges. We see the same phenomenon here, but only when the global $U(1)$ gauge symmetry remains intact. An important and interesting question for future work is precisely how the superselection seen here relates to the \textit{flux superselection} sectors studied in \autocite{gomesQuasilocalDegreesFreedom2021,rielloHamiltonianGaugeTheory2024}. Since flux superselection sectors relate to the presence of global gauge symmetries in classical field theories, it seems plausible that, through quantization, they give rise to DHR superselection sectors. This relation is studied from the perturbative point of view in \autocite{Rejzner:2020xid}.

In situation \textbf{(B)}, the global $U(1)$ symmetry is assumed to be broken by an order parameter, labelled $F_\text{SSB}$, in the sense of Proposition \ref{orderparameterprop}. In that case, everything turns out quite the opposite way. There cannot be massless contributions to the electromagnetic spectrum (as these would restore the $U(1)$ symmetry), so all bosons must be massive. This is of course the desired and usual consequence of the Higgs mechanism, and it is now formulated entirely in terms of the quantum-theoretical definition of SSB. In addition, instead of superselection we have current charge screening: the Noether relation fails and the improved smeared current $j_0(f_R\alpha_{\delta R})$ annihilates any Coulomb state $\Psi\in\mathcal{F}_C\Psi_0$ in the appropriate limit. As we mentioned before, we can think of this as being due to the impossibility of detecting charges from infinitely far away, because of the short range of the screened electromagnetic field. Lastly, the usual avoidance of massless Goldstone modes is also guaranteed.

Now, the above result is powerful, but of course we must ask ourselves to what extent it can be generalized to the non-Abelian case. Whereas there does not seem to be any direct generalization, De Palma and Strocchi have in fact proven the avoidance of Goldstone's theorem for the general case of global gauge symmetry breaking in Yang-Mills theory \autocite{depalmaNonperturbativeArgumentNonabelian2013}, as follows:

\begin{theorem}
In the BRST quantization of Yang-Mills theory with structure group $G$ (with ghost fields $c^a$ and $\bar{c}^a$ and auxiliary field $B^a$), if the global gauge group $G$ is broken by the vacuum expectation value of an element $F$ of the field algebra $\mathcal{F}$, i.e.~$\langle\delta F\rangle_0\neq 0$, then the Fourier transform of the two-point function $\langle J_\mu^a(x)F\rangle_0$, where $J_\mu^a$ are the conserved Noether currents defined by
\begin{align*}
    J^a_\mu=\partial^\nu F_{\nu\mu}+\partial_\mu B^a+f^a_{bc}A^b_\mu B^c-if^a_{bc}\bar{c}^b(D_\mu c)^c,
\end{align*}
contains a $\delta(k^2)$, i.e.~there are massless Goldstone modes. However, these modes do not belong to the physical spectrum.
\end{theorem}
Thus, also in the non-Abelian case, global gauge symmetry breaking in QFT agrees with one of the main ideas behind the Higgs mechanism: the ``eating'' of Goldstone modes. Yet, as far as the authors are aware, the actual existence of massive gauge bosons in the case of broken global non-Abelian gauge symmetry has not been proved anywhere.

\section{Conclusion}

We have discussed how the global gauge transformations are singled out as the subgroup of the gauge group that is boundary-preserving but not generated by the Gauss constraint, and therefore are not {\it redundant} but have direct physical significance, because they change the electric flux, as measured ``at infinity''\endnote{There is here an analogy with diffeomorphisms in general relativity, where diffeomorphisms that generate symmetries at infinity can have a real physical effect. For a detailed discussion, see \autocite{de2017invisibility}; see also \autocite{de2017comparing,belot2018fifty}.} 
As shown in the work of Struyve, here discussed in detail, the Abelian Higgs mechanism can be reformulated as the breaking of only the global gauge symmetry, thus assuaging the worries that various philosophers have expressed about the standard narrative of the Higgs mechanism. We have also demonstrated how this viewpoint harmonizes with the dressing field method advocated by Berghofer, François and others, but applied, along the lines of the work of Gomes and Riello and the field-dependent DFM, to the {\it redundant} gauge group $\mathcal{G}^\infty_0$ of small, local gauge transformations rather than to the structure group $U(1)$ or $SU(2)$. We have also seen how the Abelian Higgs mechanism has been proved by Morchio and Strocchi to be an instance of SSB of global $U(1)$ gauge symmetry in QFT.

These ideas are supported by the physics of superconductors, which also exhibit global and not local gauge symmetry breaking \autocite{vanwezelSpontaneousSymmetryBreaking2008}. Thus, our results suggest that the analogy between the Higgs mechanism and superconducitvity is not merely formal, as argued in \autocite{fraserHiggsMechanismSuperconductivity2016}, but physical, since global gauge symmetries are physical. 

Still, we lack any causal, dynamical account of the Higgs mechanism, which should detail precisely how global gauge symmetries are broken in some kind of early universe phase transition or crossover. However, this is the case for all SSB in quantum systems, although there have been various attempts at an account of dynamical symmetry breaking \autocite{landsmanFleaSchrodingerCat2013,landsmanFoundationsQuantumTheory2017,vandevenClassicalLimitSpontaneous2022,vanwezelPhaseTransitionsManifestation2022}. Thus, we have reduced the problem of Abelian gauge symmetry breaking to the general problem of quantum SSB. The non-Abelian case remains to be better understood, though significant steps have been made.\endnote{See \autocite{lusannaDiracObservablesHiggs1997,lusannaDiracObservablesHiggs1997a,lusannaDiracObservablesSu1998,gomesUnifiedGeometricFramework2019,gomesQuasilocalDegreesFreedom2021,rielloHamiltonianGaugeTheory2024,borsboom}} Besides extension to the non-Abelian case, another possibility for future research is to deepen the link of our work to edge modes and (quantum) reference frames, along the lines of, e.g. \autocite{donnellyLocalSubsystemsGauge2016,carrozzaEdgeModesReference2022,fewsterQuantumReferenceFrames2024,Fewster:2025ijg}.

On the philosophical side, it would be of interest to consider the implications of this work for the ontology of fibre bundles \autocite{JACOBS202334,gomesGaugeTheoryGeometrisation2025}. In particular, the idea that the global $U(1)$ gauge symmetry of electromagnetism is physical raises the question of whether a vector-bundle point-of-view (VB-PoV), as proposed in \autocite{Gomes:2025zak}, is tenable. We suspect that, in as far as the VB-PoV entails the notion that structure groups of principal bundles must be viewed as structure-preserving frame transformation groups of vector bundles, our results do not conflict with the VB-PoV. This is because part of the structure that a vector bundle frame transformation must preserve are the boundary conditions on sections of said bundle. Unbroken and broken Higgs phases respectively correspond to boundary conditions in which the Higgs field approaches zero or a fixed constant minimum. In the former case all global $U(1)$ frame transformations preserve the boundary condition, but in the latter case none of them do. Thus it seems that global $U(1)$ symmetry breaking can work with the VB-PoV.

\section*{Acknowledgements}

We thank Hessel Posthuma for collaboration on some of the topics of this paper, Klaas Landsman for providing generous feedback and guidance and Jasper van Wezel for giving detailed feedback on an earlier manuscript. We also thank two anonymous referees for their insightful comments, which greatly improved the manuscript. SB wants to thank Jeremy Butterfield for advice and discussions in Oxford, and Ward Struyve for many hours of discussion in Leuven. SDH also thanks Nicholas Teh for discussions during the early stages of this work. This work is partly supported by the Spinoza Grant of the Dutch Science Organization (NWO) awarded to Klaas Landsman.

\printendnotes

\printbibliography


\end{document}